\RequirePackage[l2tabu, orthodox]{nag}
\RequirePackage{snapshot}

\documentclass[9pt,onecolumn]{extarticle}

\makeatletter
\if@twocolumn
  \usepackage[dvips,letterpaper,top=0.5in, bottom=0.5in, left=0.75in, right=0.5in,includefoot,heightrounded]{geometry}
\else
  \usepackage[dvips,letterpaper,margin=1in,includefoot,heightrounded]{geometry}
\fi

\usepackage{srcltx}

\usepackage[russian,portuges,english]{babel}

\iflanguage{portuges}
    {\newcommand{\keywordname}{Palavras-chaves}}
    {\newcommand{\keywordname}{Keywords}}

\usepackage{amssymb,amsmath,amsfonts}
\usepackage{graphicx,rotating}
\usepackage{verbatim}
\usepackage{enumerate,graphics}
\usepackage{multirow}
\usepackage{url}
\usepackage{bm}
\usepackage[dvipsnames]{xcolor}
\usepackage{tikz}
\usepackage{wasysym}
\usetikzlibrary{arrows, automata}
\usepackage{algorithmic}
\usepackage{algorithm}
\usepackage{stackrel}
\usepackage{icomma}
\usepackage{fullpage}
\usepackage{dsfont}
\usepackage{lipsum}
\usepackage{mathtools}
\usepackage{cuted}
\usepackage{setspace}
\usepackage{subfigure}
\usepackage{booktabs}
\usepackage{appendix}
\usepackage{microtype}
\usepackage{natbib}

\usepackage{abstract}

\usepackage{flushend}

\usepackage{hyperref}

\usepackage{enumerate}

\usepackage{listings}

\usepackage[short,12hr]{datetime}
       \usepackage{fouriernc}

\newcommand*{\doi}[1]{\href{http://dx.doi.org/#1}{doi: #1}}

\makeatletter

\newcommand{\printtitle}{%
\makeatletter
\if@twocolumn

\twocolumn[%
  \maketitle
  \begin{onecolabstract}
    \myabstract
  \end{onecolabstract}
  \begin{center}
    \small
    \textbf{\keywordname}
    \\\medskip
    \mykeywords
  \end{center}
  \bigskip
]
\saythanks
\else
  \maketitle
  \begin{onecolabstract}
    \myabstract
  \end{onecolabstract}
  \begin{center}
    \small
    \textbf{\keywordname}
    \\\medskip
    \mykeywords
  \end{center}
  \bigskip
  \onehalfspacing
\fi
\makeatother
}

\author{%
B. G. Palm%
\thanks{Department of Mathematics and Natural Sciences,
	Blekinge Institute of Technology, Sweden;
Programa de P\'os-gradua\c{c}\~ao em Estat\'istica, Universidade Federal Pernambuco, Brazil;
Signal Processing Group, Departamento de Estat\'{\i}stica,
	Universidade Federal Pernambuco, Brazil.
E-mail: \url{bruna.palm@bth.se}}
\and
F. M. Bayer%
\thanks{Departamento de Estat\'{\i}stica and LACESM,
	Universidade Federal de Santa Maria, Brazil.
E-mail: \url{bayer@ufsm.br}}
\and
R. J. Cintra%
\thanks{Signal Processing Group, Departamento de Estat\'{\i}stica,
	Universidade Federal Pernambuco, Brazil;
School of Science and Mathematics,
	Howard Payne University, Texas, USA.
E-mail: \url{rjdsc@de.ufpe.br}}
}

\title{%
2-D Rayleigh Autoregressive Moving Average Model for SAR Image Modeling}

\newcommand{\myabstract}{%
Two-dimensional~(2-D)
autoregressive moving average~(ARMA) models
are
commonly applied to
describe
real-world image data,
usually
assuming
Gaussian
or
symmetric noise.
However,
real-world
data often
present
non-Gaussian
signals,
with
asymmetrical
distributions
and
strictly
positive
values.
In particular,
SAR images are known
to be well
characterized
by the Rayleigh distribution.
In this context,
the ARMA model tailored for 2-D~Rayleigh-distributed data is introduced---the
2-D~RARMA model.
The~2-D RARMA model is derived and conditional likelihood
inferences are discussed.
The proposed model
was submitted
to extensive Monte Carlo simulations
to
evaluate the
performance of the
conditional maximum likelihood
estimators.
Moreover,
in the context of
SAR image processing,
two
comprehensive numerical
experiments
were performed
comparing
anomaly
detection
and
image
modeling
results of the proposed model with traditional
2-D~ARMA models
and
competing methods
in the literature.
}

\newcommand{\mykeywords}{%
	Anomaly detection,
	ARMA modeling,
	Rayleigh distribution,
	SAR images,
	two-dimensional models
}

\date{}

\begin{document}

\printtitle

\section{Introduction}

The
parametric representation
of
two-dimensional
homogeneous random fields
considering
two-dimensional (2-D)
autoregressive moving average~(ARMA) models
is frequently
adopted
for image
processing \citep{Bustos2009a,
	kizilkaya2005,
	ojeda2010new,
	bustos2009},
including
(i)~modeling~\citep{Bustos2009a,rosenfeld2014};
(ii)~compression~\citep{nijim1996};
(iii)~encoding~\citep{chung19922};
and
(iv)~restoration~\citep{lim1990,vallejos2004}.
The
2-D~ARMA model
is an spatial extension
of
the classical one-dimensional
(1-D)
ARMA model~\citep{brockwell2016},
and
it
is often employed
in
edge detection~\citep{ojeda2010new}
and
stochastic
texture analysis~\citep{hall1995}.

In particular,
the two-dimensional autoregressive first-order model
is commonly applied to
describe
real-world image data~\citep{kashyap1988,bennett1999},
representing
different types of textures~\citep{bustos2009}.
Theoretical details
of the
two-dimensional autoregressive first-order model,
such as
properties, correlation structure, and maximum likelihood
estimators of the parameters
can be found in~\cite{basu1993}.
The
ARMA model
is
preferable over
the autoregressive~(AR)
and moving average~(MA) model,
since it provides
more effective models
for
homogeneous random fields~\citep{cadzow1981,zhang1991}.
In fact, ARMA models
can better
characterize
the power in the spectral
domain representation
when
compared to
AR or MA models~\citep{kizilkaya2005}.

ARMA modeling
usually
assumes Gaussian
or
symmetric noise
distribution,
incurring inferential problems when this assumption
is not satisfied~\citep{kizilkaya2005,zoubir2018,schwartz1989topics}.
Indeed,
the Gaussianity
hypothesis
has been widely
considered in statistical
signal processing~\citep{zoubir2018},
remote sensing analysis~\citep{zhao2008,
	zhao2016,
	morales2017},
and
detection theory~\citep{
	kay2000,
	xue2020,
	wang2019}.
However,
actual measured signals often
present
non-Gaussian properties~\citep{
	margoosian2015,
	liu2018},
such as
asymmetrical distributions
and
strictly
positive
values.

An alternative approach
for modeling
such type of data
is the use of the
Rayleigh distribution~\citep{Bayer2019b}.
This
model
is considered
in
signal
and image
processing~\citep{
	zanetti2015,
	sumaiya2018},
being relevant
in the context of
synthetic aperture radar
(SAR) image
modeling,
due to its
good characterization
of image pixel amplitude
values~\citep{yue2019generalized,
	Kuruoglu2004,
	Jackson2009,kuttikkad2000statistical}.
In particular,
this distribution
is known to well fit
SAR data homogeneous regions~\citep{oliver2004,Yue2021}.
The main justification for supposing
the Rayleigh distribution as a model for
the amplitude of SAR data is that the assumption
of a large number of reflectors in an observed image
allows one to invoke the central limit theorem,
according to which the distribution of the real
and complex parts
of the received signal are independent
and approaches the Gaussian distribution
with zero mean and
constant variance~\citep{Kuruoglu2004,Yue2021}.
Thus,
under these assumptions,
the amplitude values of complex SAR data are exactly Rayleigh distributed.

Frequently,
the SAR image modeling is performed assuming
constant parameters~\citep{Kuruoglu2004,Jackson2009}.
In cases where this assumption is not suitable,
an
alternative is to use a regression model,
where each observation has one specific
estimated mean~\citep{wang2008,McCullagh1989,palm2019}.
Motivated by
these SAR image characteristics,~\cite{palm2019}
proposed
a regression model
based on the Rayleigh distribution
for SAR image modeling.
However,
image
pixels
usually
present a resolution spatial dependence~\citep{Jackson2009,yan2018real},
and
consequently,
a~2-D model
can be used as a venue
for addressing such a problem.

The use of one- or two-dimensional dynamical models
are often employed in image applications.
For example,
in~\cite{almeida2021}, the authors
proposed a new
1-D ARMA model
considering the~$\mathcal{G}_0$ distribution
to estimate the intensity values of SAR image pixels.
\cite{Bayer2019b} derived
an 1-D Rayleigh-based dynamical model
useful to land-use classification in SAR images.
In~\cite{bustos2009},
a 2-D Gaussian ARMA model
was applied in image filtering schemes.
Additionally, the literature
presents several studies
for~1-D dynamical models
based on different distributions,
useful in a multitude of scientific applications;
see.,
e.g.,~\cite{benjamin2003,bayer2017,rocha2009,moller2020generalized,Palm2021}.

However,
to the best of our knowledge,
a two-dimensional ARMA model
assuming the Rayleigh distribution
is not present in the literature
and this paper aims at proposing a first treatment
on the topic.
Our goal is two-fold.
First,
we derive
a two-dimensional~ARMA model
for
non-Gaussian
spatial autocorrelated images,
where the observed signal
is
asymmetric
and
strictly
positive.
For the proposed
model,
we
introduce
parameter
estimation,
large data record inference,
and
the quantile residuals.
Second,
we propose
an
image
modeling tool
based on the
derived spatial model estimated parameters
and an
anomaly
detector
for
non-Gaussian SAR images.
Considering
control
charts
of the proposed model
residuals,
the introduced detection scheme
measures the deviations of an observed pixel value
from its estimated mean value.
The residual-based control charts
have already been
employed
in remote
sensing data
change
and anomaly
detection~\citep{bayer20203,brooks2013}.

Anomaly detection
is a popular field in
signal processing,
machine learning, and
statistics~\citep{talagala2020,kadri2016,quatrini2020,kwon2005}.
In particular,
anomaly detection in noisy image
is explored
for quality control purposes
in
different manufacturing applications,
such as composites,
steel,
and textile production~\citep{yan2018real}.
The anomaly detection problem
can be addressed
considering different techniques,
depending on the way
that anomalies are defined.
For example,
different types of
expected outputs
or input data
have particular
problem formulation
and need to be
addressed through specific
techniques~\citep{talagala2020}.
On the other hand,
our proposal
is based on a simple
residual analyze
considering
control chart with a fixed theoretical threshold,
which is
defined based on the residual distribution.

This paper is organized as follows.
In Section~\ref{s:model_2d}, we describe
the proposed spatial model,
provide
conditional
maximum likelihood
estimators,
and present a
hypothesis testing
methodology.
Section~\ref{s:changedetector}
details
the introduced
image modeling
tool
and
proposed
an
anomaly
detection algorithm.
Section~\ref{s:num_2d}
presents
Monte Carlo
simulations
and
two
empirical
analyses
of the derived
detector
applied to SAR images.
Section~\ref{s:conclu_2d}
brings
final remarks
and concludes
the paper.

\section{The Proposed Model}
\label{s:model_2d}

\subsection{Mathematical Setup}

Recently,
a regression model~\citep{palm2019}
and
an 1-D~ARMA model~\citep{Bayer2019b}
based on the Rayleigh distribution
have been proposed.
The Rayleigh ARMA~(RARMA)
model
introduced in~\cite{Bayer2019b}
relates
the mean
of
an
one-dimensional
discrete-time signal
to a linear predictor
through a
strictly monotonic,
twice differentiable
link function~$g(\cdot)$,
where
$g:\mathbb{R}^+\!\rightarrow\mathbb{R} $.
The goal of this
section is to
extend
the 1-D
model
presented in~\cite{Bayer2019b}
and
introduce
to the~2-D case.

Let~$Y[n,m]$,
$n = 1,2,\ldots , N$,
$m = 1,2, \ldots , M$,
be
a
random
variable
representing
the pixels
of an~$N \times M$
image;
and let~$y[n,m]$
be
the realization
of the
signal~$Y[n,m]$.
Additionally,
let~$S [n,m] =
\lbrace [k,l] \in  \mathds{Z}^2 :
1 \leq
k \leq n,
1 \leq
l \leq m \rbrace
-
\lbrace [n,m] \rbrace$
be
the strongly causal region at~$[n,m]$~\citep{bustos2009}.
Assume that,
conditionally to the
information
set in the neighborhood~$S[n,m]$,
each~$Y[n,m]$ is distributed according to
the
Rayleigh distribution.
Considering the
mean-based parametrization
of~$Y[n,m]$
propose in~\cite{palm2019},
we have that
the conditional density of~$Y[n,m]$,
given $S [n,m] $,
is provided by
\begin{align*}
f_Y(y[n,m]\mid S [n,m] )
=&
\frac{\pi y[n,m] }{2 \mu[n,m]  ^2}
\exp\left(-\frac{\pi y[n,m] ^2}{4 \mu[n,m] ^2}\right)
,
\end{align*}
where~$\mu[n,m] > 0 $.
The cumulative distribution function is given by
\begin{align*}
F_Y(y[n,m] \mid S [n,m])
= 1- \exp\left(-\frac{\pi y[n,m]^2}{4 \mu[n,m]^2}\right).
\end{align*}
The conditional mean and conditional
variance of~$Y[n,m]$,
given~$S [n,m]$,
are,
respectively,
expressed by means of
\begin{align*}
\begin{split}
\operatorname{E}(Y[n,m]\mid S [n,m])
&= \mu[n,m],
\\
\operatorname{Var}(Y[n,m] \mid S [n,m])
&=
\mu[n,m] ^2 \left(\frac{4}{\pi}-1 \right)
.
\end{split}
\end{align*}

\subsection{The Model}

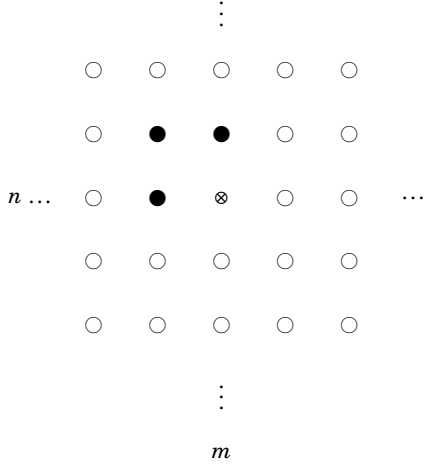
\begin{figure}
	\centering
	\begin{tikzpicture}[scale=0.85]
	\draw
	(3, 3) node[circle, black](a0){$\vdots$}
	(1, 2) node[circle, black](a1){$\Circle$}
	(2, 2) node[circle, black](a2){$\Circle$}
	(3, 2) node[circle, black](a3){$\Circle$}
	(4, 2) node[circle, black](a4){$\Circle$}
	(5, 2) node[circle, black](a5){$\Circle$}
	(1, 1) node[circle, black](a1){$\Circle$}
	(2, 1) node[circle, black](a2){$\newmoon$}
	(3, 1) node[circle, black](a3){$\newmoon$}
	(4, 1) node[circle, black](a4){$\Circle$}
	(5, 1) node[circle, black](a5){$\Circle$}
	(0, 0) node[circle, black](a0){$n$ $\ldots$}
	(1, 0) node[circle, black](a1){$\Circle$}
	(2, 0) node[circle, black](a2){$\newmoon$}
	(3, 0) node[circle, black](a3){$\otimes$}
	(4, 0) node[circle, black](a5){$\Circle$}
	(5, 0) node[circle, black](a5){$\Circle$}
	(6, 0) node[circle, black](a0){$\ldots$}
	(1, -1) node[circle, black](a1){$\Circle$}
	(2, -1) node[circle, black](a2){$\Circle$}
	(3, -1) node[circle, black](a3){$\Circle$}
	(4, -1) node[circle, black](a4){$\Circle$}
	(5, -1) node[circle, black](a5){$\Circle$}
	(1, -2) node[circle, black](a1){$\Circle$}
	(2, -2) node[circle, black](a2){$\Circle$}
	(3, -2) node[circle, black](a3){$\Circle$}
	(4, -2) node[circle, black](a4){$\Circle$}
	(5, -2) node[circle, black](a5){$\Circle$}
	(3, -3) node[circle, black](a0){$\vdots$}
	(3, -4) node[circle, black](a0){$m$};
	\end{tikzpicture}

	\caption{Example of
		the
		neighborhood
		used
		in
		a 2-D~RARMA$(1,1)$
		model.
	}

	\label{f:2dex1}

\end{figure}

The proposed 2-D~Rayleigh
autoregressive
and moving average model,
hereafter referred
to as~2-D RARMA,
is defined
according to
\begin{align}
\label{e:edmodelo2}
\Phi{(z_1,z_2)}  g(y[n,m])
=
\beta +
\Theta{(z_1,z_2)}  e[n,m]
,
\end{align}
where
the
two-dimensional autoregressive
operator~$\Phi{(z_1,z_2)} $
and moving average
operator~$\Theta{(z_1,z_2)} $
are furnished, respectively, by
\begin{align*}
\begin{split}
\Phi{(z_1,z_2)}
=&
1
-
\sum \limits_{i=0}^p
\sum \limits_{j=0}^p \phi_{(i,j)}z_1^{i}z_2^{j}
,
\\
\Theta{(z_1,z_2)}
=&
1 +
\sum \limits_{k=0}^q
\sum \limits_{l=0}^q \theta_{(k,l)}z_1^{k}z_2^{l}
,
\end{split}
\end{align*}
and
$  \beta \in \mathbb{R} $
is
an
intercept;
the quantities
$z_1^{i}g(y[n,m]) =  g(y[n-i,m])$,
$z_2^{j}g(y[n,m]) =  g(y[n,m-j])$,
$z_1^{k}e[n,m] = e[n-k,m]$,
and
$z_2^{l}e[n,m] =  e[n, m-l]$
are the
backward operators;
$p$
and~$q$ are the orders of the model;
the quantities~$\phi_{(i,j)}$,~$i,j=0,1,\ldots,p$,
and~$\theta_{(k,l)}$,~$k,l=0,1,\ldots,q$,
are,
respectively,
the autoregressive and moving average parameters
estimated
based on the
image pixels;
$e[n,m] =
g(y[n,m]) - g(\mu[n,m])$
is the moving average error term;
and
$g(\mu[n,m] )= \eta[n,m]$
the
linear predictor.
As suggested
in~\cite{basu1993}
for an unilateral spatial ARMA,
we assume
$\phi_ {(0,0)}
= \theta_{(0,0)}
= 0$.
Replacing
the quantities
described above
in~\eqref{e:edmodelo2},
the 2-D~RARMA~($p,q$) model
can be rewritten as
\begin{align}
\label{e:modeldef}
\begin{split}
g(\mu [n,m])
=&
\beta
+
\sum \limits_{i=0}^{p}
\sum \limits_{j=0}^{p}
\phi_{(i,j)}
g(y [n-i,m-j])
+
\sum \limits_{k=0}^{q}
\sum \limits_{l=0}^{q}
\theta_{(k,l)}
e[n-k,m-l]
.
\end{split}
\end{align}
Figure~\ref{f:2dex1}
depicts
the
considered
neighbors
pixels
in a
2-D~RARMA$(1,1)$
model.

As
in 1-D non-Gaussian dynamical models
considering
conditional probability models under different distributions,
conditions of weak stationarity, causality,
and invertibility
are still
challenging and
open topics in the literature.
This limitation is related to the fact
that,
for link functions other than the identity,
the MA error terms
are not martingale
differences,
which make the
first two moments of the marginal distribution
intractable~\citep{benjamin2003}.
However, this fact does not make
such models less useful in practice,
being widely used in
different areas of knowledge,
see, e.g.,~\cite{melchior2021,almeida2021,
	leiva2020,rothermel2020,liboschik2017,scher2020}.

\subsection{Conditional Likelihood Estimation}
\label{s:estimation_2d}

The estimation of
the
2-D RARMA$(p,q)$ model parameters
can be
realized by maximizing the
logarithm of the conditional
likelihood function~\citep{brockwell2016}.
Let
$\bm{\gamma}=(\beta,
\bm{\phi}^\top,\bm{\theta}^\top)^\top$
be
the parameter vector
where
$\bm{\phi} = (\phi_{(0,1)}, \phi_{(0,2)},
\ldots, \phi_{(p,p)})^\top$
and
$\bm{\theta} = (\theta_{(0,1)}, \theta_{(0,2)},
\ldots, \theta_{(q,q)})^\top$,
with dimensions~$(p+1)^2-1$
and~$(q+1)^2-1$,
respectively.
The
log-likelihood function for~$\bm{\gamma}$,
conditional to
the~$w =\max(p,q)$ preliminary observations,
is given by
\begin{align*}
\ell(\bm{\gamma})=&
\sum_{n=w+1}^{N}
\sum_{m=w+1}^{M}
\log f_Y(y[n,m]\mid S[n,m])
=
\sum_{n=w+1}^{N}
\sum_{m=w+1}^{M}
\ell[n,m](\mu[n,m])
,
\end{align*}
where
\begin{align*}
\begin{split}
\ell[n,m](\mu[n,m])
=&
\log\left(\frac{\pi}{2}\right) +
\log(y[n,m])
- \log(\mu[n,m]^2)
-\frac{\pi y[n,m]^2}{4 \mu[n,m]^2}
.
\end{split}
\end{align*}
The
conditional maximum likelihood estimator~(CMLE),
$\widehat{\bm{\gamma}}$,
can be obtained
by solving
\begin{align}
\label{e:escore22}
\mathbf{U}(\bm{\gamma}) =
\frac{\partial \ell}{\partial \bm{\gamma}^\top} =
\left( \frac{\partial \ell }{\partial \beta},
\frac{\partial \ell }{\partial \bm{ \phi}^\top},
\frac{\partial \ell }{\partial \bm{\theta}^\top},
\right)^\top
=
\bm{0}
,
\end{align}
where~$\mathbf{U}(\bm{\gamma})$
is the score vector
and~$\bm{0}$
is the vector of zeros
with dimension~$(p+1)^2 + (q+1)^2 -1 $.
Computing the derivatives
in~\eqref{e:escore22},
we obtain
\begin{align*}
\frac{\partial \ell}{\partial
	\bm{\gamma}} =&
\sum \limits _{n=w+1} ^{N}
\sum \limits _{m=w+1} ^{M}
\frac{\partial \ell[n,m](\mu[n,m]) }{\partial \mu[n,m]}
\frac{\operatorname{ d }  \mu[n,m]}{\operatorname{ d } \eta[n,m]}
\frac{\partial \eta[n,m]}{\partial
	\bm{\gamma}}
.
\end{align*}
Note that
$
\frac{\operatorname{ d }  \ell[n,m](\mu[n,m])}{\operatorname{ d }  \mu[n,m]}
=
\frac{\pi y[n,m]^2}{2\mu[n,m]^3}-\frac{2}{\mu[n,m]}
$
and
$
\frac{\operatorname{ d }  \mu[n,m]}{ \operatorname{ d }  \eta[n,m]}
= \frac{1}{g'(\mu[n,m])}
$,
where~$g'(\cdot)$ is the first derivative of
the selected link function~$g(\cdot)$.
Thus,
we can write
\begin{align*}
\begin{split}
\frac{\partial \ell}{\partial
	\bm{\gamma}} =&
\sum \limits _{n=w+1} ^{N}
\sum \limits _{m=w+1} ^{M}
\left(
\frac{\pi y[n,m]^2}{2\mu[n,m]^3}-\frac{2}{\mu[n,m]}
\right)
\frac{1}{g'(\mu[n,m])}
\frac{\partial \eta[n,m]}{\partial
	\bm{\gamma}}
,
\end{split}
\end{align*}
where
\begin{align*}
\begin{split}
\frac{\partial \eta [n,m]}{\partial \beta}
=&
1
-
\sum \limits _{s=0}^{q}
\sum \limits _{t=0}^{q}
\theta_{(s,t)}
\frac{\partial \eta[n-s,m-t]}{\partial \beta}
,
\\
\frac{\partial \eta [n,m]}{\partial  \phi_{(i,j)}}
=&
g(y[n-i,m-j])
-
\sum \limits _{s=0}^{q}
\sum \limits _{t=0}^{q}
\theta_{(s,t)}
\frac{\partial \eta [n-s,m-t]}
{\partial \phi_{(i,j)}}
,
\\
\frac{\partial \eta [n,m]}{\partial  \theta_{(k,l)}}
=&
g(y [n-k,m-l])
- g(\mu[n-k,m-l])
-
\sum \limits _{s=0}^{q}
\sum \limits _{t=0}^{q}
\theta_{(s,t)}
\frac{\partial \eta[n-s,m-t]}
{\partial \theta_{(k,l)}}
,
\end{split}
\end{align*}
for~$(i,j) \in
\lbrace 0,1,\ldots,p \rbrace^2 -
\lbrace (0,0) \rbrace$
and~$(l,k)
\in
\lbrace  0,1,\ldots,q \rbrace^2 -
\lbrace (0,0) \rbrace$ .

The
Broyden-Fletcher-Goldfarb-Shanno~(BFGS)
method~\citep{press}
with analytic first derivatives
was adopted
as the
nonlinear
optimization algorithm~\citep{nocedal1999}
to solve~\eqref{e:escore22}.
The~BFGS method was selected
due to its
superior performance for
non-linear
optimization~\citep{Mittelhammer2000}.
The
initial values for
the constant~($\beta$)
and
the autoregressive~($\bm{\phi}$) parameters
were
derived
from the ordinary least squares estimate
associated to the linear regression
model,
which has a closed matrix form.
The response vector
is
$(g(y[w+1,w+1]), g(y[w+1,w+2]),
\ldots, g(y[N,M]))^\top$
and
the covariate matrix is given by
	\begin{align*}
	\begin{bmatrix}
	1
	& g(y[w,w-1]) &
	g(y[w-1,w])
	& \cdots &
	g(y[w-p, w-p]) \\
	1
	& g(y[w,w])
	& g(y[w,w])
	& \cdots
	& g(y[w-p+1, w-p+1]) \\
	\vdots
	& \vdots
	& \vdots
	& \ddots & \vdots  \\
	1
	& g(y[N,M-1])
	& g(y[N-1,M])
	& \cdots &
	g(y[N-p , M- p])
	\end{bmatrix}.
	\end{align*}
As
suggested in~\cite{Bayer2019b},
we set~$\bm{\theta} = \mathbf{0}$
as initial values.

\subsection{Large Data Record Inference}
\label{s:wald}

Under some usual regularity conditions,
the CMLE
is consistent and
asymptotically normally distributed~\citep{andersen1970asymptotic}.
Thus, in large data records,
we have that
\begin{align*}
(\widehat{\bm{\gamma}}-\bm{\gamma})
\stackrel[{N \cdot M \rightarrow \infty}]{d}{\longrightarrow}
\mathcal{\mathbf{N}} \left(\bm{0},
\mathbf{I}^{-1}(\bm{\gamma})\right)
,
\end{align*}
where~$\stackrel{d}{\longrightarrow} $
represents convergence in distribution
and
$\mathcal{\mathbf{N}}( \bm{0} , \mathbf{I}^{-1}(\bm{\gamma}))$
denotes the
multivariate
Gaussian distribution
with null mean
and
covariance matrix~$\mathbf{I}^{-1}(\bm{\gamma})$.
The
conditional Fisher information matrix,~$\mathbf{I}(\bm{{\gamma}})$,
is discussed in detail
in the Appendix.

To derive a
hypothesis testing methodology
tailored for the
2-D~RARMA model parameters,
the likelihood-based detection theory~\citep{Pawitan2001,Kay1998-2}
can be considered.
Let~$\bm{\gamma}$
be
partitioned
in a parameter vector
of interest~$\bm{\gamma}_I$
of dimension~$\nu$,
and
a vector of nuisance parameters~$\bm{\gamma}_J$
of dimension~$[(p+1)^2 + (q+1)^2 -1 ]
- \nu$~\citep{Kay1998-2}.
In addition,~$\mathcal{H}_0:\bm{\gamma}_{I}=\bm{\gamma}_{I_0}$
is the hypothesis of interest
and~$\mathcal{H}_1:\bm{\gamma}_{I} \neq \bm{\gamma}_{I_0}$
the alternative hypothesis,
where~$\bm{\gamma}_{I_0}$
is a fixed column vector of dimension~$\nu$.
The
Wald statistic
is given by~\citep{Kay1998-2}
\begin{align}
\label{e:wald}
T_W
&=
(\widehat{\bm{\gamma}}_{I_1}-\bm{\gamma}_{I_0})^\top
\left( \left[ \mathbf{I} ^ {-1}
(\widehat{\bm{\gamma}}_1) \right] _{\gamma_I \gamma _I} \right)^{-1}
(\widehat{\bm{\gamma}}_{I_1}-\bm{\gamma}_{I_0})
,
\end{align}
where
$\widehat{\bm{\gamma}}_1
= (\widehat{\bm{\gamma}}_{I_1}^\top,
\widehat{\bm{\gamma}}_{J_1}^\top)^\top$
is the CMLE under~$\mathcal{H}_1$
and
$
\left[ \mathbf{I}^{-1} (\widehat{\bm{\gamma}}_1)
\right]_{\gamma_I \gamma _I}
$
is a
partition
limited to the
estimates of interest
of
the
estimated
conditional Fisher information matrix.
Under~$\mathcal{H}_0$,
the test statistic,~$T_W$,
asymptotically
follows
the
chi-square distribution
with~$\nu$
degrees of freedom,
$\chi^2_\nu$~\citep{Kay1998-2}.
The hypothesis test
consists of
comparing the computed value
of~$T_W$
with
a threshold value,~$\epsilon$,
which
is
obtained
based on
the~$\chi^2_\nu$ distribution
and
the
desired
probability of false alarm~\citep{Kay1998-2}.

To test the
overall significance of
a fitted
model,
we considered
the following hypotheses
\begin{align}
\label{e:hip_apli}
\begin{cases}
\mathcal{H}_0 : \bm{\gamma}^\star= \mathbf{0}
,
\\
\mathcal{H}_1 : \bm{\gamma}^\star \neq \mathbf{0}
,
\end{cases}
\end{align}
where~$\bm{\gamma}^\star = (\bm{\phi}^\top, \bm{\theta}^\top)^\top$.
Using
the
Wald test
described above,
we
reject~$\mathcal{H}_0$
when~$T_W > \epsilon$.
In this situation,~$\bm{\gamma}^\star \neq \mathbf{0}$,
indicating that at least some of
the autoregressive and moving average
parameters
are nonzero
and the spatial correlation among the pixels
is significant.
Additionally,
a
confidence interval~(CI)
for the $i$th component
of~$\bm{\gamma}$,
$i = 1 , 2, \ldots , (p+1)^2 + (q+1)^2 -1$,
with confidence approximately~$100(1-\alpha)\%$,
can be derived based on
the CMLE asymptotic distribution
as
\begin{align*}
\left[\widehat{\gamma}_i - z_{1-\alpha/2}
\text{se}(\widehat{\gamma}_i);
\widehat{\gamma}_i + z_{1-\alpha/2}
\text{se}(\widehat{\gamma}_i)\right],
\end{align*}
where
$\text{se}(\widehat{\gamma}_i) = \sqrt{\mathbf{I}^{-1}_{ii}(\widehat{\bm{\gamma}})}$
is the standard error (SE) of $\widehat{\gamma}_i$,
$\mathbf{I}_{ii}^{-1}(\widehat{\bm{\gamma}})$
is the~$i$th element of the
diagonal of~$\mathbf{I} ^{-1}(\widehat{\bm{\gamma}})$,~$\alpha$
is the significance level,
and~$z_{\varrho}$
is
the~$\varrho$th quantile of the
standard normal distribution.

\section{Image Modeling and Anomaly Detection}
\label{s:changedetector}

In this section,
we propose
an image modeling
and
an anomaly
detection
tool
based on
the proposed
2-D~RARMA model.
We also discuss
model selection strategies.
For such,
we
introduce
the estimated values of~$\mu [n,m]$
and
present the residuals
of
the
2-D~RARMA model.

\subsection{Image Modeling}

The modeled image
is obtained by
applying
the estimated values
of~$\mu [n,m]$,~$\widehat{\mu}[n,m]$,
in the
2-D~RARMA$(p,q)$
model structure,
given by~\eqref{e:edmodelo2},
and
evaluating
it
at~$\bm{\widehat{\gamma}}$.
Therefore,
the fitted signal
is given by
\begin{align}
\label{e:filter2d}
\nonumber
\widehat{\mu} [n,m]
=&
g^{-1}
\big{(}
\widehat{\beta}
+
\sum \limits_{i=0}^{p}
\sum \limits_{j=0}^{p}
\widehat{\phi}_{(i,j)}
g(y [n-i,m-j])
+
\sum \limits_{k=0}^{q}
\sum \limits_{l=0}^{q}
\widehat{\theta}_{(k,l)}
e [n-k,m-l]
\big{)}
,
\end{align}
where~$n = w +1, w+2, \ldots , N$
and~$m = w +1, w+2, \ldots ,M$.
Hence,
similar to the 1-D model,
the image border
is not included
in the modeling process,
since
the resulting
fitted image
has~$(N-w) \times (M-w)$
pixels.

\subsection{Anomaly Detector}
\label{s:anomalydetec}

Residuals
can be useful for
performing a
diagnostic analysis of the fitted model
and
can be defined as
a function of the
observed
and
predicted values~\citep{kedem2005}.
We employ
the
quantile
residuals~\citep{Dunn1996},
defined
as
\begin{align*}
r[n,m] &=
\Phi^{-1}\left(
F_Y
(
y[n,m] \mid S [n,m]
)
\right)
,
\end{align*}
where
$\Phi^{-1}$ denotes the standard normal
quantile function.
When
the model is correctly fitted,
for construction,
the quantile residuals are
approximately
Gaussian distributed
with zero mean and
unit
variance,
i.e.,~$r[n,m]  \sim \mathcal{N} (0,1)$~\citep{Dunn1996}.
The residual analysis for other classes of
1-D non-Gaussian dynamical models is
discussed in~\cite{kedem2005}
and~\cite{Dunn1996}.

Large values of
$r[n,m]$
can be interpreted as
anomaly
changes
in the image
behavior.
To capture such variations
of the residual values,
we adopted the use of control charts.
Since the residuals
have
approximately
unitary variance,
the control chart
detects
an image change
if the residual value
is outside the control limit~$\pm L$.
We adopted~$L=3$,
since
it is expected that residuals are
randomly distributed around zero
and
inside the interval~$[-3,3]$,
about~$99.7\%$
of the observations,
since
$
2\Phi (L) - 1
\vert_{L=3}
\approx
99.7 \%
$~\citep{brooks2013,bayer2019}.
If the residual value
is outside
this range,
the analyzed pixel
is understood
to differ
from the
expected
behavior
according to the
2-D~RARMA model
fitted in
the
region
of interest
and,
consequently,
some
anomaly
change
might have occurred.

Notice that
the proposed model
relies
on neighboring pixels
from the northwest direction,
as shown in Figure~\ref{f:2dex1}.
Thus,
to take into account the other directions
in an
omnidirectional
manner,
thus ensuring that all surrounding pixels
are considered,
the 2-D~RARMA fitting is also
applied
to the~$90^{\circ}$,~$180^{\circ}$,
and~$270^{\circ}$
rotated region of interest
to capture information from
the
versions of the
southwest, southeast, and northeast directions.
Results are combined according to the morphological
union of the resulting binary images.
Search for the best model for each rotation
is computationally expensive.
Consequently,
considering a trade-off between
simplicity and efficacy of the anomaly detection method,
we suggest using the same model order
for the four directions.

To further increase the performance
of the proposed
detector,
a post-processing
step using
mathematical
morphological operations,
such as
erosion,
dilation,
opening,
and
closing
operations,
can be considered~\citep{edmond2000mathematical,gonzales2009}.
Such operations
aim at
(i)~removing
small
spurious pixel groups
which are regarded as noise
and
(ii)~preventing
the
splitting
of the interest targets
into multiple
substructures~\citep{gonzales2009}.
The resulting data
is the detected image.
The proposed
ground type
change detection method
is summarized in
Algorithm~\ref{a:alg1}.

\begin{algorithm}
	\centering
	\caption{
		Anomaly
		detection
		method
		based on the 2-D~RARMA$(p,q)$ model
	}
	\label{a:alg1}
	\begin{algorithmic}
		\REQUIRE
		Interest image $\mathbf{X}_{\text{input}}$

		\ENSURE
		Anomaly detection
		image $\mathbf{X}_{\text{detected}} $
		\STATE 1) Select region of interest
		$\mathbf{X}_{\text{selected}} \subset \mathbf{X}_{\text{input}}$
		which
		anomaly detection
		is to be tested against.
		\STATE 2)
		Fit the 2-D~RARMA$(p,q)$ model
		for the
		following images:

		$\quad \mathbf{X}_0 = \mathbf{X}_{\text{selected}}$

		$\quad \mathbf{X}_k = \textbf{rot90}(\mathbf{X}_{k-1})$,

		for~$k = 1,2,3$,
		where
		$\textbf{rot90}(\cdot)$
		rotates its argument
		counterclockwise
		by 90 degrees.
		\STATE 3)
		For each resulting fitted image,
		compute
		residuals~$r_k[n,m]$
		relative to
		$\mathbf{X}_{\text{input}}$.
		\STATE 4)
		Obtain four binary images as follows
		\IF{$(r_k[n,m]  \leq -3) \,\,\, \text{or} \,\,\, (r_k[n,m]  \geq 3)$}
		\STATE $ \tilde{X}_k[n,m] \leftarrow 1 $
		\ELSE
		\STATE $\tilde{X}_k[n,m] \leftarrow 0$
		\ENDIF

		for~$k = 0,1,2,3$.
		\STATE 5)
		Compute binary image from the following pixel-wise Boolean union:
		$\mathbf{\tilde{X}}
		\leftarrow
		\bigcup\limits_{k=0}^{3}
		\mathbf{\tilde{X}}_k
		$.
		\STATE 6) Apply morphological operators as a final post-processing step:
		$\mathbf{X}_{\text{detected}}
		\leftarrow
		\text{post-processing}(\mathbf{\tilde{X}})$.

	\end{algorithmic}
\end{algorithm}

\subsection{Model Selection}
\label{s:modelselection}

To perform the model selection,
we considered the
three-stage iterative Box-Jenkins methodology~\citep{Box2008},
which is based on the following steps:
identification, estimation,
and statistical model checking.
For the first step,
we suggest to use the
Akaike's~(AIC)~\citep{Akaike}
and
Schwartz's~(SIC)~\citep{schwarz1978}
information criteria
to define
some rough boundaries on the choice of~$p$ and~$q$ orders.
The~AIC and~SIC are given by
\begin{align*}
\text{AIC} =&
- 2 \ell (\bm{\widehat{\gamma}})
+
2 \kappa
,
\\
\text{SIC} =&
- 2 \ell (\bm{\widehat{\gamma}})
+ \kappa \log (N \times M)
,
\end{align*}
where~$\kappa = (p+1)^2 + (q+1)^2 -1$ is the number
of estimated parameters in the fitted model.
Lower AIC and SIC values
are related to more suitable models.
The estimation step is done by
the conditional maximum likelihood method,
as explored in Section~\ref{s:estimation_2d}.

Regarding the model checking step,
we suggest
to consider the following approach:
(i)~test the overall significance of the fitted model
through the Wald test introduced in Section~\ref{s:wald};
(ii)~perform graphic analysis
to check whether the model residuals are
randomly distributed around zero;
and
(iii)~compute the mean square
error~(MSE)
and
the
mean absolute percentage error~(MAPE)
between~$y[n,m]$ and~$\widehat{\mu}[n,m]$
among the candidate fitted models.
Lower MSE and MAPE values
are specially desired in detection problems
because
good quality predictions
imply in accurate detection results.
As well as in the Box-Jenkins method,
if the fitted model is not suitable
according to any of the steps described above,
then
we have to return
to the first step and attempt to build a better model.

\section{Numerical Results}
\label{s:num_2d}

In this section,
we
aim at
evaluating the CMLE of
the
2-D~RARMA model
parameters
and
assessing
the
performance
of the proposed
image modeling and
anomaly
detector.
For such,
the proposed
analyses were
performed in the context
of SAR image processing.
We performed three
numerical
experiments:
(i)~a simulated data analysis
to assess the proposed estimators
and
(ii)~two
computations aiming at
anomaly
detection
based on
actual SAR images.

\subsection{Analysis with Simulated Data}
\label{s:simu}

Rayleigh distributed data~$y[n,m]$
were generated
by the
inversion method
with mean given by~\eqref{e:edmodelo2}
and
logarithm link function.
We
considered
simulations
under
two
scenarios:
(i)~2-D~RARMA$(1,0)$ model
and
(ii)~2-D~RARMA$(1,1)$ model.
The
parameter values
were selected
based on
estimated
values
of the
2-D~RARMA
model parameters
from a
SAR image
forest region
acquired by
the airbone CARABAS~II system~\citep{Lundberg2006},
a
Swedish ultrawideband (UWB)
very-high frequency~(VHF) SAR device
operating at horizontal~(HH) polarization.
Details related to the data
can be found in~\cite{
	Ulander2005}
and~\cite{Lundberg2006}.
The
obtained
numerical values of the
parameters
were~$\beta= -0.2031$,
$\phi _{(0,1)} = 0.4562  $,
$\phi _{(1,0) }= 0.4523   $,
and~$\phi _{(1,1) }= -0.1054$,
for the 2-D~RARMA$(1,0)$ model;
and
$\beta= 0.3569$,
$\phi _{(0,1)} = 0.2155$,
$\phi _{(1,0)} = 0.2032$,
$\phi _{(1,1)} = 0.1500$,
$\theta _{(0,1)} = 0.1529$,
$\theta _{(1,0)} = 0.1744$,
and~$\theta _{(1,1)} = 0.1998$,
for the 2-D~RARMA$(1,1)$ model.
The number of Monte Carlo replications
was set to~$1,000$ and the
sizes of the
synthetic images
were~$ \{
10 \times 10,
20 \times 20 , 40 \times 40 ,
80 \times 80
\}$.

\begin{figure}
	\centering
		\subfigure[Total Relative Bias]
		{\label{f:bias}\includegraphics[width=0.48\textwidth]{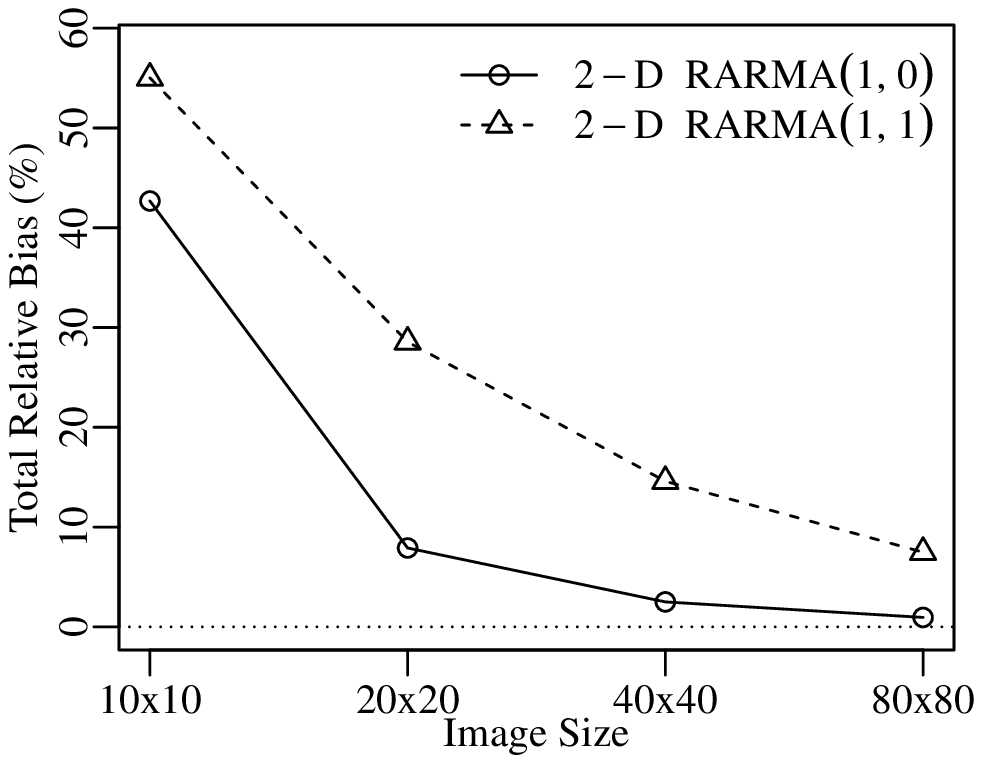}}
		\subfigure[Total MSE]
		{\label{f:mse}\includegraphics[width=0.48\textwidth]{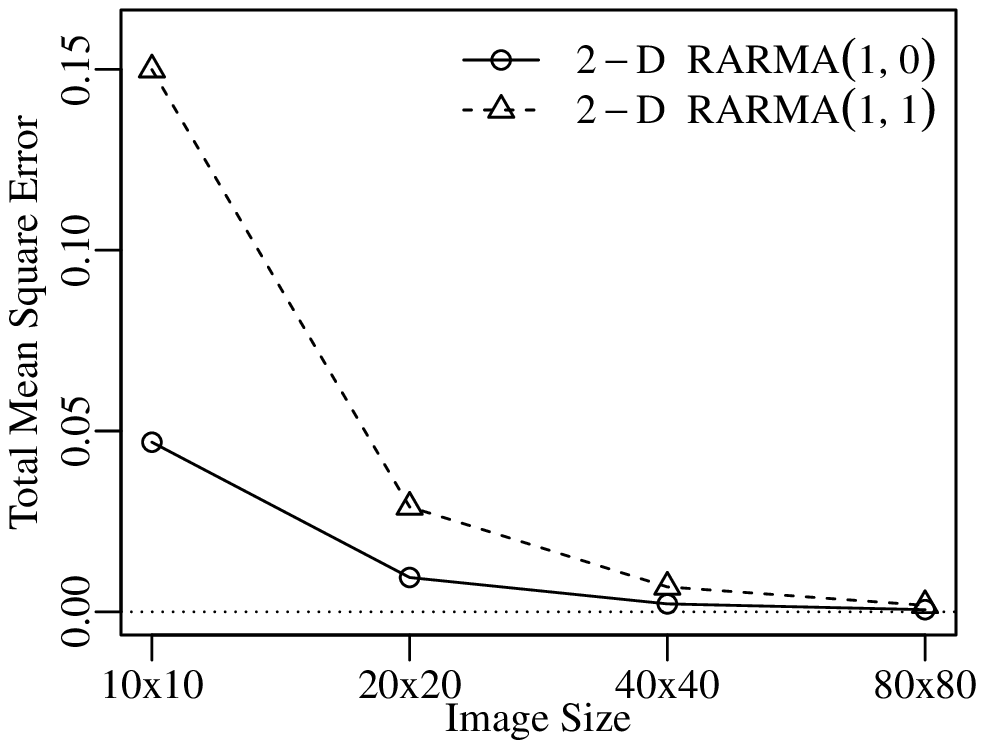}}
		\caption{
			Total relative bias and mean square error of the
			2-D RARMA($1,0$) and 2-D RARMA($1,1$) model
		estimators.}
\end{figure}

In order to numerically evaluate the point estimators,
we computed
the
mean,
percentage relative bias~(RB\%),
and
mean square error~(MSE)
of the CMLE.
A graphical analysis of the
point estimators
is performed through
graphs of total relative bias
and total mean square error.
Both measures are defined
as the sum of the absolute values
of the individual
relative biases and mean square error
and are displayed in Figures~\ref{f:bias}
and~\ref{f:mse}, respectively.

Tables~\ref{t:rar1}
and~\ref{t:rarma}
present the simulation results
for 2-D~RARMA$(1,0)$
and
2-D~RARMA$(1,1)$
models,
respectively.
As expected,
both
bias and MSE
figures
improve
when larger images are considered.
This behavior
is in agreement with the
asymptotic property
(consistency)
of the CMLE.
Convergence failures were absent
for all considered scenarios.
In contrast
to the
traditional
2-D~ARMA model,
the proposed model
avoids
the
estimation
problem
of the MA parameters,
as discussed in~\cite{kizilkaya2005}
and~\cite{lim1990}.
The
derived model
estimates
the
AR and MA terms
simultaneously;
and ~$\bm{\widehat{\phi}}$
and~$\bm{\widehat{\theta}}$
present
closer
values
of
RB(\%)
for all considered
synthetic images.
The image size of~$40 \times 40$
was
sufficiently large
for accurate inference
in
2-D~RARMA$(1,0)$
model,
i.e.,
MSE and RB(\%) values
close to zero.
On the other hand,
the
2-D~RARMA$(1,1)$
model
shows accurate inference
results
for an
image size of $80 \times 80$
pixels.
We can verify graphically
from Figures~\ref{f:bias}
and~\ref{f:mse}
that the RB\% and the MSE
decrease as the image size increases in
both evaluated models,
numerically indicating the estimator consistency.

To evaluate the interval estimators,
we computed
the coverage rate~(CR)
of the confidence intervals
with a significance level of $5\%$.
The~CR is defined according the
following steps:
(i)~in each Monte Carlo replication,
compute the~CI
and
interrogate whether the~CI contains the true parameter
or not
and
(ii)~define the percentage
of replications
for which
the
parameter
is
in the~CI.
It is
desirable that the CR
approaches the nominal coverage level~$100(1-\alpha)\%$.
The
CR values
in
Tables~\ref{t:rar1}
and~\ref{t:rarma}
are close to the nominal value
of~$0.95$
in all considered scenarios,
specially for larger image sizes.

\begin{table}
	\centering
	\caption{
		Simulation results on point
		and interval
		estimation
		of the 2-D~RARMA$(1,0)$ model
	}
	\label{t:rar1}
	\centering
	\begin{tabular}{lcccc}
		\toprule
		Measures 	&
		Mean &
		RB(\%)    &
		MSE &
		CR\\
		\midrule
		\multicolumn{5}{c}{ $N=M =10$ } \\
		\midrule
		$\widehat{\beta}$ 	&	$-0.2629$ &   $29.4370$ &    $0.0244$ &
		$0.9230$ \\
		$\widehat{\phi}_{(0,1)}$ & $ 0.4385$ &  $ -3.8767$  &   $0.0071$ &  $0.9400$ \\
		$\widehat{\phi}_{(1,0)}$ &   $0.4339$ &   $-4.0692$ &    $0.0073$ &  $0.9360$ \\
		$\widehat{\phi}_{(1,1)}$ &    $-0.0998$ &   $-5.2979$ &   $ 0.0081$  &  $0.9430$	\\
		\midrule
		\multicolumn{5}{c}{ $N=M =20$ } \\
		\midrule
		$\widehat{\beta}$ 	&  $-0.2160$ &   $ 6.3468$ &    $0.0051$ &
		 $0.9360$	 \\
		$\widehat{\phi}_{(0,1)}$ & $0.4523$ &   $-0.8514$ &    $0.0014$ &
		  $0.9400$\\
		$\widehat{\phi}_{(1,0)}$ & $ 0.4492$ &  $ -0.6841$ &   $ 0.0014$ &
		  $0.9510$ \\
		$\widehat{\phi}_{(1,1)}$   &	$-0.1054$ &   $-0.0223$ &   $ 0.0016$ &
		 $0.9600$ \\
		\midrule
		\multicolumn{5}{c}{ $N=M =40$ } \\
		\midrule
		$\widehat{\beta}$ 	&	$-0.2063$ &   $ 1.5865 $ &   $0.0012$ &
		 $0.9520$ \\
		$\widehat{\phi}_{(0,1)}$ &  $ 0.4551$ &  $ -0.2409$ &   $ 0.0003$ &
		 $0.9550$ \\
		$\widehat{\phi}_{(1,0)}$ &  $ 0.4514$ &   $-0.2021$ &   $ 0.0003$ &
		 $0.9440$ \\
		$\widehat{\phi}_{(1,1)}$  &  $-0.1049$ &   $-0.4706$ &   $ 0.0004$ &
		 $0.9520$ \\
		\midrule
		\multicolumn{5}{c}{ $N=M =80$ } \\
		\midrule
		$\widehat{\beta}$ 	&	$-0.2043$ &    $0.6020$ &   $ 0.0003$ &
		 $0.9520$\\
		$\widehat{\phi}_{(0,1)}$ & $ 0.4560$ &  $ -0.0464$ &    $0.0001$ &
		$0.9430$\\
		$\widehat{\phi}_{(1,0)}$ &  $0.4516$ &   $-0.1504$ &   $ 0.0001$ &
		 $0.9520$\\
		$\widehat{\phi}_{(1,1)}$  & $-0.1052$ &   $-0.1450$ &   $ 0.0001$ &
		  $0.9640$\\
		\bottomrule
	\end{tabular}
\end{table}

\begin{table}
	\centering
	\caption{
		Simulation results on point
		and interval
		estimation
		of the 2-D~RARMA$(1,1)$ model
	}
	\label{t:rarma}
	\centering
	\begin{tabular}{lcccc}
		\toprule
		Measures 	&
		Mean &
		RB(\%)    &
		MSE &
		CR\\
		\midrule
		\multicolumn{5}{c}{ $N=M =10$ } \\
		\midrule
		$\widehat{\beta}$ 	& $0.2834$ &  $-20.5927$ &  $0.0223$ &
		$0.7960$ \\
		$\widehat{\phi}_{(0,1)}$ & $0.2278$   &$ 5.7302$  &  $0.0405$ &
		 $0.8400$ \\
		$\widehat{\phi}_{(1,0)}$ &  $0.2266$  & $11.5074$  &  $0.0388$ &
	$0.8180$\\
		$\widehat{\phi}_{(1,1)}$   &	 $0.1758$  & $17.1830$  &  $0.0483$ &   $0.7680$ \\
		$\widehat{\theta}_{(0,1)}$ & $0.1371$  &$-10.3477$  &  $0.0583$ &
		 $0.8100$ \\
		$\widehat{\theta}_{(1,0)}$ & $0.1413$  &$-18.9945$  &  $0.0604$ &
		$ 0.8060$\\
		$\widehat{\theta}_{(1,1)}$   & $0.1582$  &$-20.8355$  &  $0.0493$ &
		 $0.8350$ \\
		\midrule
		\multicolumn{5}{c}{ $N=M =20$ } \\
		\midrule
		$\widehat{\beta}$ 	&	$0.3312$&   $-7.2026$    &$0.0036$ &
		 $0.9060$\\
		$\widehat{\phi}_{(0,1)}$ &   $0.2246$&    $4.2177$    &$0.0078$ & $0.9140$ \\
		$\widehat{\phi}_{(1,0)}$ &   $0.2124$&    $4.5210$    &$0.0081$ &
		 $0.9120$ \\
		$\widehat{\phi}_{(1,1)}$   &	 $0.1689$&   $12.6150$    &$0.0095$ &  $0.8890$ \\
		$\widehat{\theta}_{(0,1)}$ &  $0.1428$&   $-6.5797$    &$0.0090$ &
		$0.9190$ \\
		$\widehat{\theta}_{(1,0)}$ & $0.1673$&   $-4.0788$    &$0.0089$ &
		$0.9120$ \\
		$\widehat{\theta}_{(1,1)}$   & $0.1753$&  $-12.2650$    &$0.0061$ &
		 $0.9200$ \\
		\midrule
		\multicolumn{5}{c}{ $N=M =40$ } \\
		\midrule
		$\widehat{\beta}$ 	&	$0.3456$&   $-3.1658$&    $0.0009$ &
		 $0.9200$ \\
		$\widehat{\phi}_{(0,1)}$ & $0.2195$&    $1.8514$&    $0.0018$ &
		 $0.9400$ \\
		$\widehat{\phi}_{(1,0)}$ &  $0.2077$&    $2.2177$&    $0.0019$ &
		 $0.9290$ \\
		$\widehat{\phi}_{(1,1)}$   &	$0.1610$&    $7.3560$&    $0.0023$ &
		 $0.9150$ \\
		$\widehat{\theta}_{(0,1)}$ &   $0.1499$&   $-1.9723$&    $0.0020$ &
		 $0.9390$ \\
		$\widehat{\theta}_{(1,0)}$ &  $0.1711$&   $-1.9205$&    $0.0020$ &
	 $0.9310$ \\
		$\widehat{\theta}_{(1,1)}$   & $0.1878$&   $-5.9962$&    $0.0013$ &	   $0.9280$ \\
		\midrule
		\multicolumn{5}{c}{ $N=M =80$ } \\
		\midrule
		$\widehat{\beta}$ 	&	 $0.3514$ &   $-1.5381$  &  $ 0.0002$ &
		 $0.9100$ \\
		$\widehat{\phi}_{(0,1)}$ &  $0.2184  $&  $1.3555  $&  $0.0005$ &
		$0.9290$ \\
		$\widehat{\phi}_{(1,0)}$ & $0.2059  $&  $1.3082  $&  $0.0005$ &
		 $0.9400$ \\
		$\widehat{\phi}_{(1,1)}$   &	$0.1549 $&   $3.2791 $&   $0.0006$ &
		 $0.9360$ \\
		$\widehat{\theta}_{(0,1)}$ &   $0.1503 $&  $-1.7221 $&   $0.0005$ &
	 $0.9420$ \\
		$\widehat{\theta}_{(1,0)}$ &  $0.1717 $&  $-1.5643 $&   $0.0005$  &
		 $0.9330$ \\
		$\widehat{\theta}_{(1,1)}$   & 	$0.1932$&   $-3.3146$&    $0.0003$ &  $0.9350$ \\
		\bottomrule
	\end{tabular}
\end{table}

\subsection{Analysis with Real Data }

In this section,
we perform two anomaly detection
experiments
considering two different amplitude SAR images.
For that, we use the methodology proposed in Section~\ref{s:anomalydetec}.
We do not have access to the complex data
of the tested SAR images,
and consequently,
it is not possible to check
if
the real
and complex parts
approaches the Gaussian distribution
with zero mean and
constant variance, as discussed in~\citep{Kuruoglu2004,Yue2021}.
However,
evidence of the usefulness of the Rayleigh distribution
for describing image pixel amplitude values
is found in~\cite{Kuruoglu2004,Yue2021}.
Moreover, the results obtained
in this paper suggest that the 2-D RARMA model
provides good anomaly detection
and modeling for SAR data.

\subsubsection{CARABAS~II}
\label{s:carabas}

\begin{figure}
	\centering
	\includegraphics[scale=0.38]{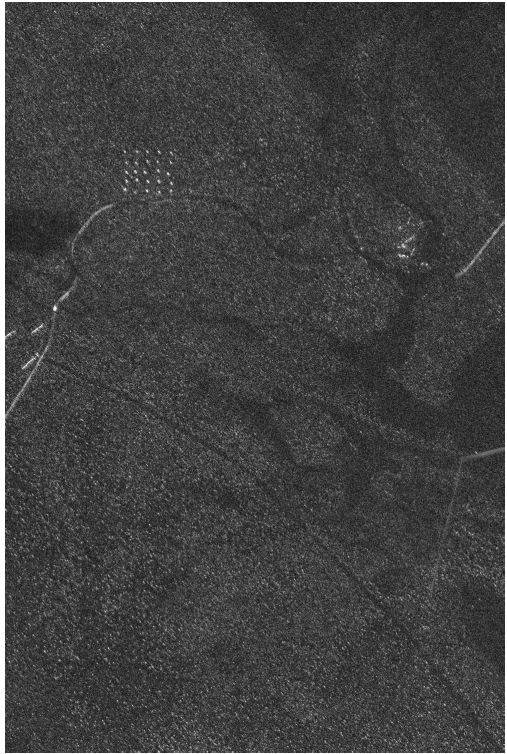}
	\caption{Original CARABAS II SAR image. }
	\label{f:carabas_1}
\end{figure}

The SAR image considered in this
experiment
was collected by the CARABAS~II system
described in the previous subsection.
As reported in~\cite{Ulander2005},
the  spatial resolution of CARABAS~II is
about
$3~\text{m}$ in azimuth
and
range.
The ground scene
of
the selected image
is
dominated by pine forest,
a lake,
and~$25$ military vehicles~\citep{Lundberg2006}.
The forest and
lake regions
characterize
most of the image area
and
they
follow a homogeneous pattern.
The military vehicles
deployed in the SAR scene~\citep{Lundberg2006}
introduce more
representative
behavior changing
when compared
to the forest and
lake regions.
For instance,
the sample mean value
of
forest, lake and military vehicles
areas are
about~$0.1267$,~$0.1148$,
and~$0.2863$, respectively,
i.e.,
the sample mean value
of military vehicles region
is roughly three times the forest and lake regions.
Pixels related to power line areas
show similar amplitude values
with the targets,
and consequently,
are strongly associated
to the false alarm detection,
as discussed in~\cite{Lundberg2006}.
The
considered SAR image
in this experiment
is shown in Figure~\ref{f:carabas_1}.

\begin{figure}

	\centering

\subfigure[Northwest model]{
	\includegraphics[scale=0.62]{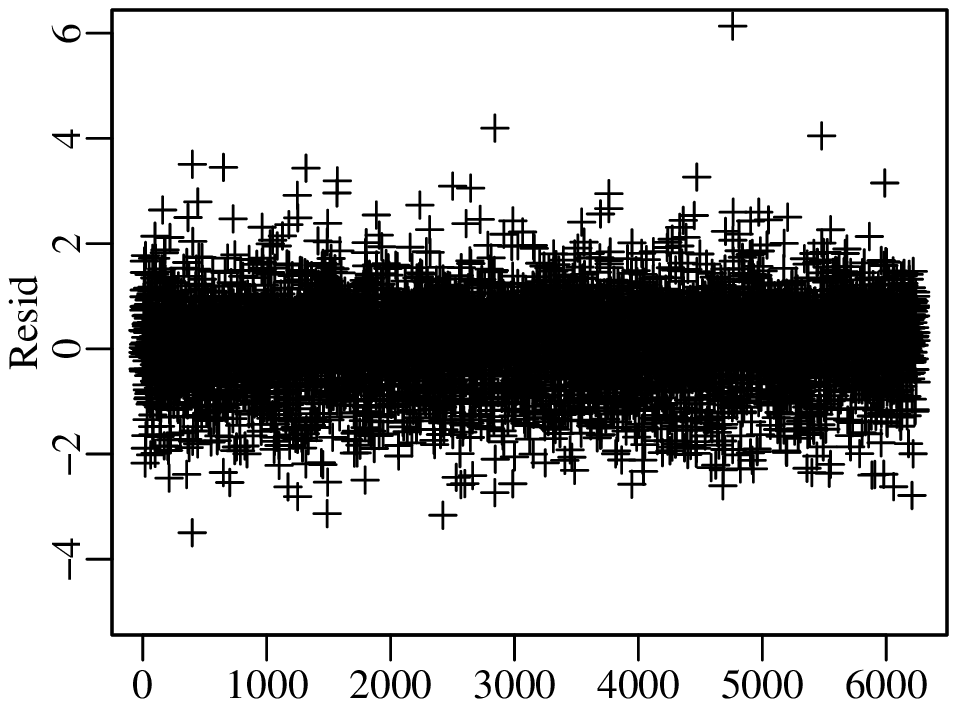}
}
\subfigure[Southwest model]{
	\includegraphics[scale=0.62]{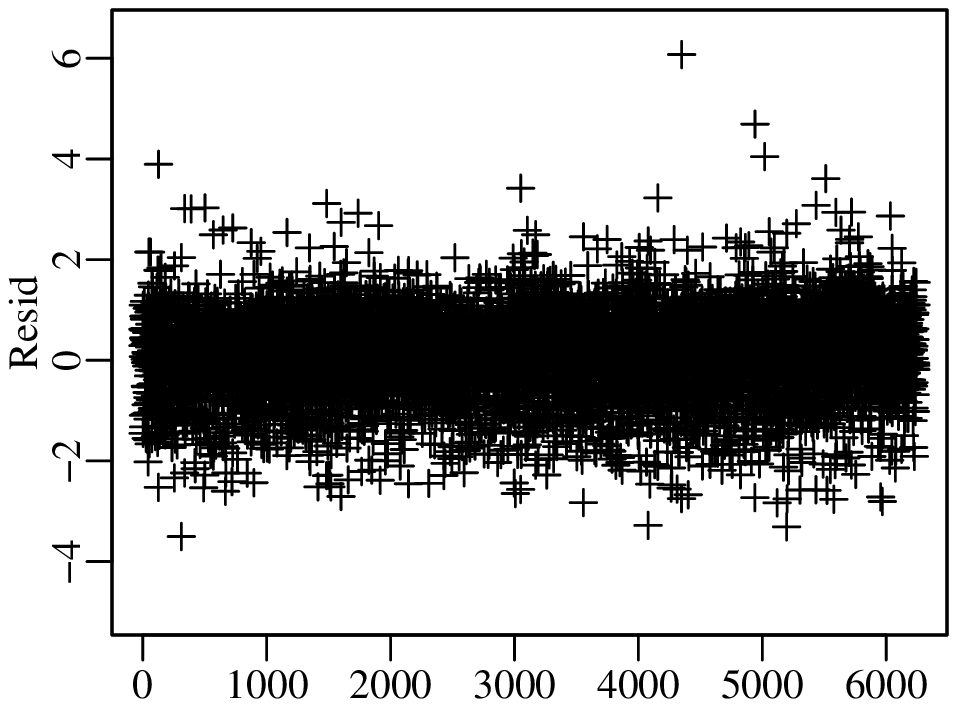}
}
\subfigure[Southeast model]{
	\includegraphics[scale=0.62]{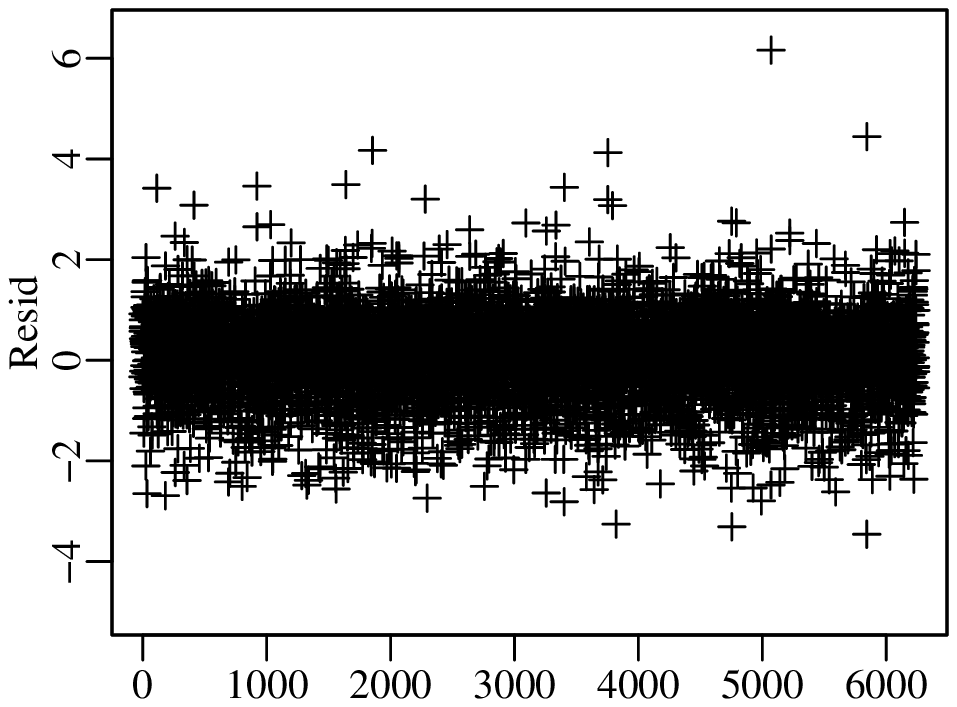}
}
\subfigure[Northeast model]{
	\includegraphics[scale=0.62]{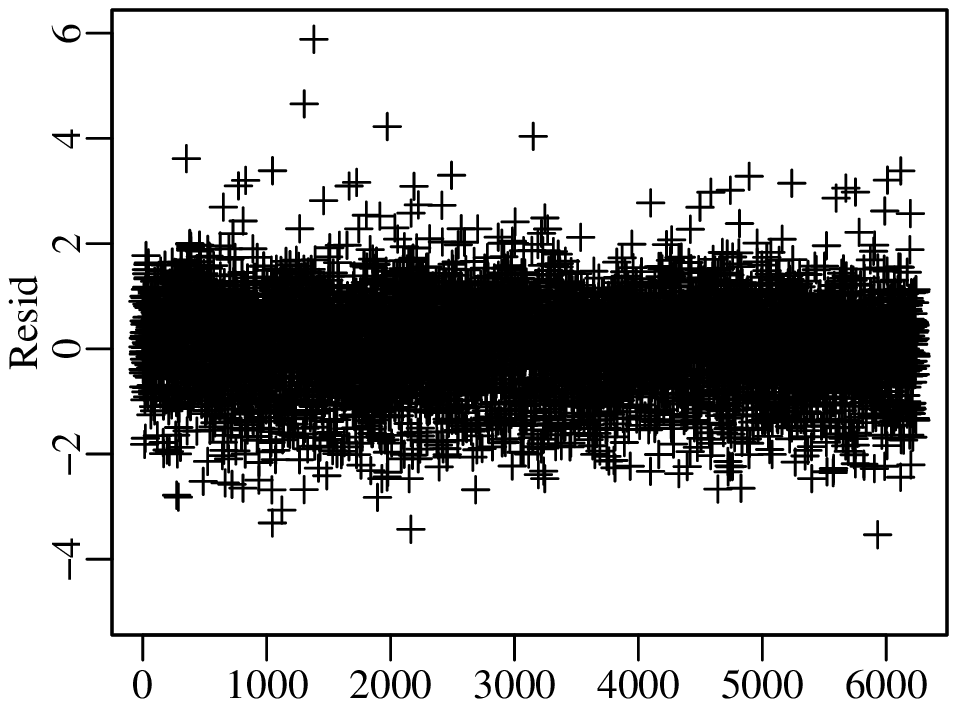}
}
	\caption{
		Residual charts of the
		2-D~RARMA$(1,1)$ models for the CARABAS~II SAR image.
	}
	\label{f:carabas_hist}
\end{figure}

The model selection
was based on the steps
described in Section~\ref{s:modelselection}.
We employed
the
search space restricted
to models with~$(p,q) \in \lbrace 0, 1, 2 \rbrace$
and
the
size of the
considered
region of interest
was
$N = M \in
\lbrace
10 ,
20 ,
40 ,
80
\rbrace$.
As a result,
we obtained
the following optimal parameters:
(i)~$p=q=1$
and
(ii)~$N=M=80$.
The selected region of interest in this section
was forest.
For the post-processing step,
we followed
the methodology
defined in~\cite{Lundberg2006},
where
we considered
two
morphology
operations,
namely,
an opening
operation
followed by a dilation.
The opening uses a
$3 \times  3~\text{pixels}$
square structuring element,
whose size is determined by the
system resolution;
the dilation
considers a
$7 \times  7~\text{pixels}$
structuring element,
which is linked
to the approximate size
of the military vehicles.
Table~\ref{t:sf_33}
shows the
estimated parameters
and standard error~(SE)
for each rotated image,
as described in
the second step
of the Algorithm~\ref{a:alg1}.
The
overall significance
Wald test
$p$-value
can be found in
Table~\ref{t:sf_33},
showing
that
the spatial autocorrelation
is significant
for a probability of false alarm
equal to $0.05$.
Figure~\ref{f:carabas_hist}
shows the residual versus index charts
of the 2-D~RARMA($1,1$) models.
As expected, the residuals
are randomly distributed
and
present values close to zero.

\begin{table*}
	\centering
	\caption{
		Estimated parameters,
		standard error~(SE),
		and $p$-values
		of the
		overall significance Wald test of the
		2-D~RARMA$(1,1)$ model
		for the CARABAS~II SAR image
		}
	\label{t:sf_33}
	\begin{tabular}{cccccccccc}
		\toprule
		\multicolumn{9}{c}{Rotated Image} \\
		\midrule
		& \multicolumn{2}{c}{ Northwest}  & \multicolumn{2}{c}{ Southwest}  &
		\multicolumn{2}{c}{ Southeast} & \multicolumn{2}{c}{ Northeast}  \\
		\midrule
		& Estimate & SE & Estimate & SE &
		Estimate &  SE & Estimate &  SE \\
		\midrule
		$\widehat{\beta}$ & $-1.2274$ &  $0.0681$ &$ -1.1146$ &  $0.0854$ &$-1.1986$ &  $0.0666$ &$ -1.2076$ & $0.0892$\\
		$\widehat{\phi}_{(0,1)}$ & $ 0.1723$ &  $0.0303$ &$ 0.1659$ & $0.0396$ &$ 0.2218$ & $0.0308$ &$ 0.1912$ & $0.0392$ \\
		$\widehat{\phi}_{(1,0)}$ & $ 0.1526$ &  $0.0316$ &$0.2206$ &  $0.0351$ &$0.1572$ & $0.0316$ &$ 0.1616$ & $0.0361$\\
		$\widehat{\phi}_{(1,1)}$  & $0.0675$ &  $0.0303$  &$ 0.0512$ & $0.0275$ &$ 0.0294$ &  $0.0308$ & $ 0.0387 $ & $0.0270$ \\
		$\widehat{\theta}_{(0,1)}$ & $ 0.1773$ & $0.0329$ &$0.1263$ &
		$0.0418$ &$0.1305$ &  $0.0336$ & $0.1127$ & $0.0416$ \\
		$\widehat{\theta}_{(1,0)}$ & $ 0.1646$ &  $0.0338$ &$ 0.1208$ &  $0.0378$ & $0.1808$ &  $0.0335$ &$0.1685$ &$0.0383$ \\
		$\widehat{\theta}_{(1,1)}$  &$ 0.1935$ & $0.0252$ &$-0.0691$ &  $0.0250$ &$0.2064$ &  $0.0251$ &$ -0.0461$ & $0.0256$  \\
		\midrule
		$p$-value & \multicolumn{2}{c}{ $<0.001$} &  \multicolumn{2}{c}{ $<0.001$}
		& \multicolumn{2}{c}{ $<0.001$} & \multicolumn{2}{c}{ $<0.001$}  \\
		\bottomrule
	\end{tabular}
\end{table*}

\begin{figure}

	\centering

	\subfigure[2-D~RARMA$(1,1)$ model]{
		\includegraphics[scale=0.3]{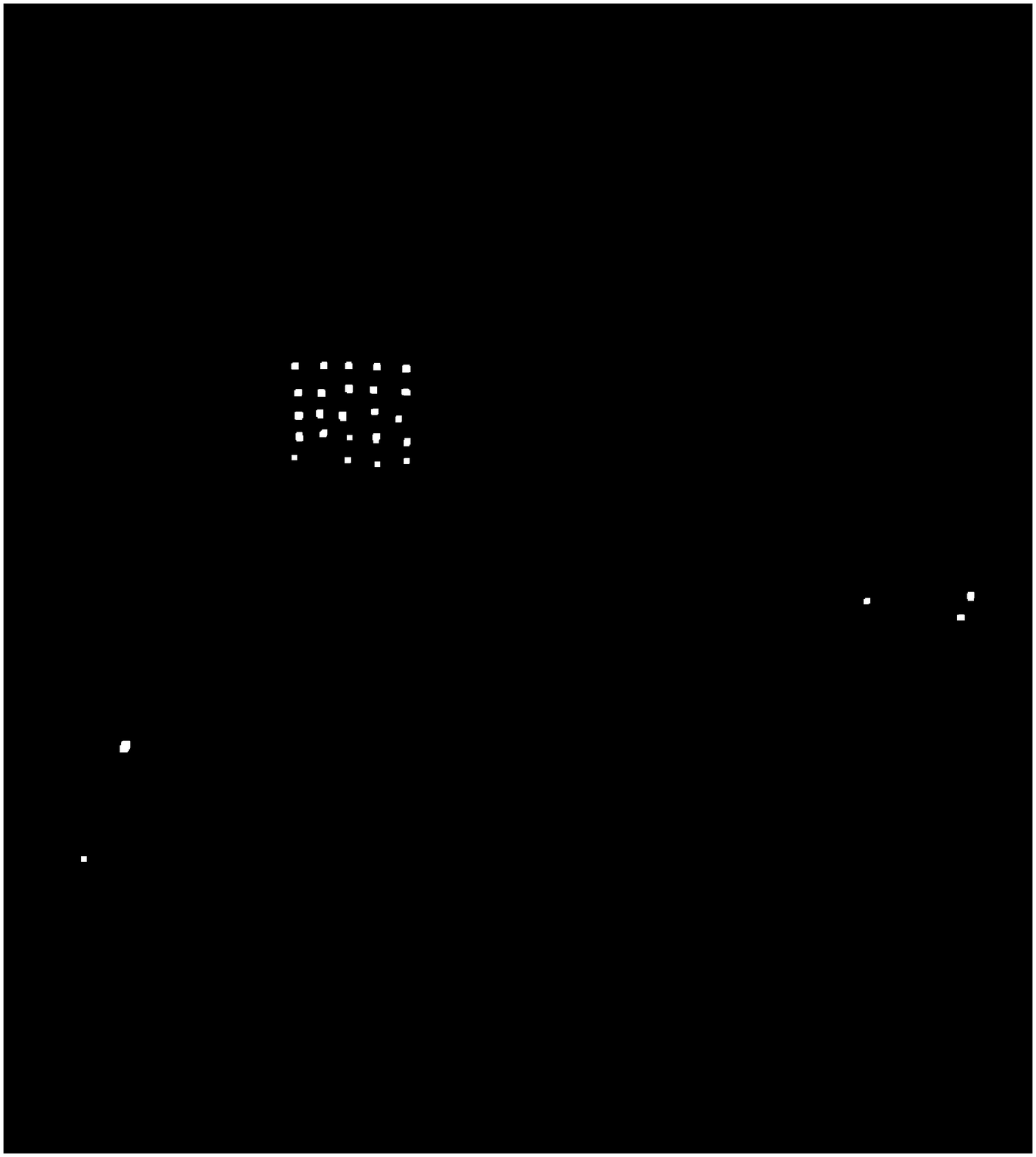}
		\label{f:carabas_rarma}
	}
	\subfigure[2-D~ARMA$(1,1)$ model]{
		\includegraphics[scale=0.3]{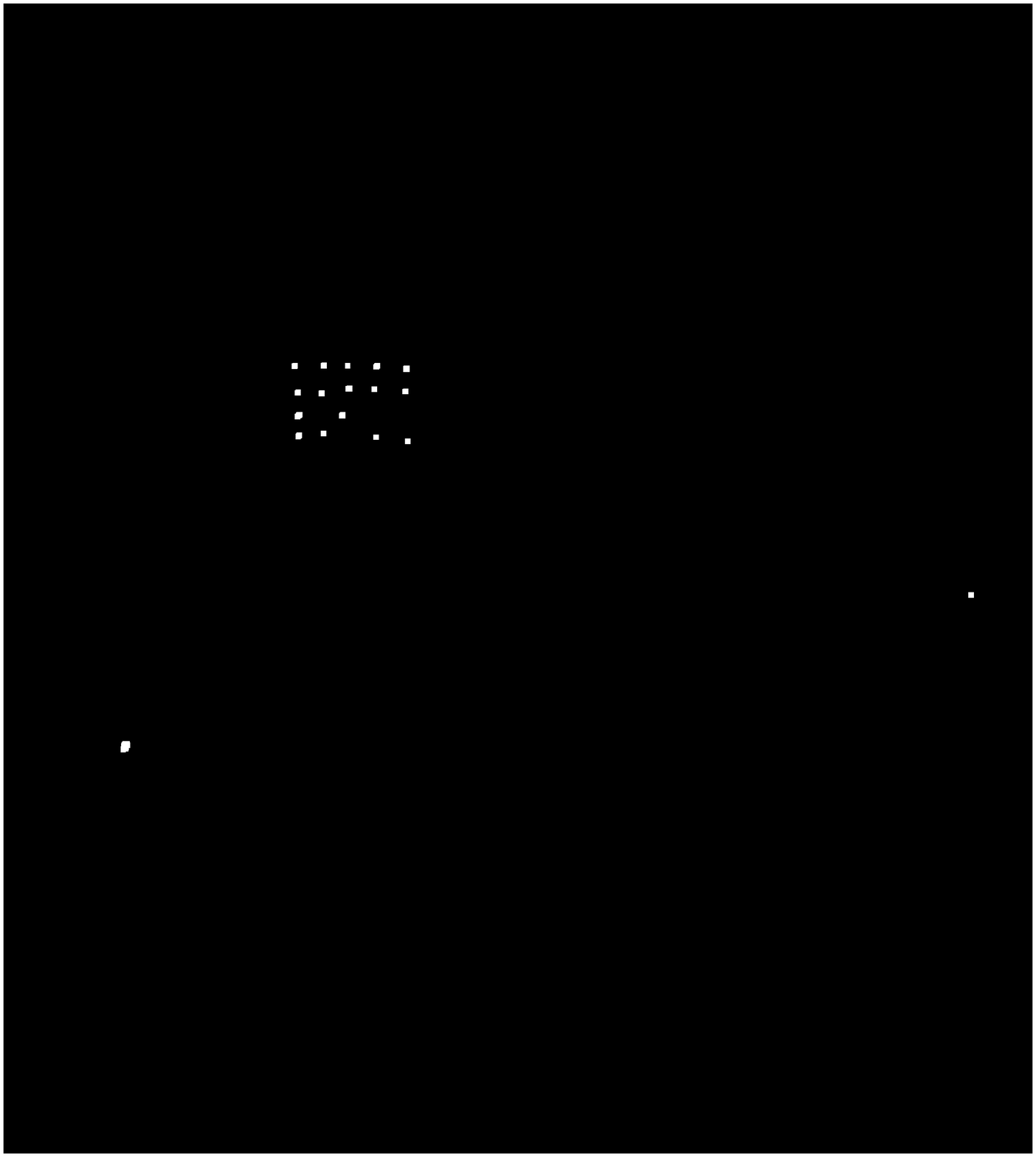}
		\label{f:carabas_arma}
	}
	\caption{
		Zoomed detected images
		in the region where
		anomaly
		changes
		were identified,
		considering
		2-D~RARMA$(1,1)$ and 2-D~ARMA$(1,1)$ models.
	}
	\label{f:sf_22}
\end{figure}

For comparison purposes,
we
also obtained the detection results
based on
the
2-D~ARMA$(1,1)$ model.
Detection results
for both
2-D~RARMA$(1,1)$
and
2-D~ARMA$(1,1)$ models
can be found in
Figures~\ref{f:carabas_rarma}
and~\ref{f:carabas_arma},
respectively.
For a better visualization,
Figure~\ref{f:sf_22}
shows the zoomed images
in the region where
ground
type changes were detected.
The proposed method
detected~$24$ military vehicles
and five false alarms.
In contrast,
the 2-D~ARMA$(1,1)$ model
can only
detect~$16$
military vehicles
and two false alarms.

We also compared the proposed methodology
with three different competing approaches:
(i)~constant false alarm rate filtering combined with
likelihood ratio test
assuming the Gaussian distribution~\citep{Lundberg2006},
(ii)~a statistical hypothesis test for wavelength-resolution
incoherent SAR
ground type
change detection
based on the
bivariate gamma
distribution~\citep{vu2018};
and
(iii)~a statistical
hypothesis test considering the
bivariate Gaussian distribution~\citep{Vu2017}.
Differently from the above methods,
our detection scheme
requires only one input image for analysis;
whereas
two
images are demanded
in~\cite{Lundberg2006} and \cite{vu2018},
respectively,
and
three
in~\cite{Vu2017}.
Despite requiring much less
assumptions and
data information
(only one image scene look
vs several image scene looks)
when compared to~\cite{Lundberg2006},~\cite{vu2018},
and \cite{Vu2017},
the
proposed
2-D~RARMA($1,1$) model
performance was very close to the competing methods:
only one less detection hit;
and 3 to 5 more false alarms.

To further
compare
the
image
modeling performance
of the evaluated models,
we
computed the MSE and MAPE
of all pixels.
Table~\ref{t:mse2} summarizes
the results
of
the quality adjustment measures
for 2-D~RARMA$(1,1)$
and
2-D~ARMA$(1,1)$
models.
The 2-D~RARMA$(1,1)$ model
excels
in terms of
MSE and MAPE
measures.
The execution time of this experiment
considering the Rayleigh- and Gaussian-based models
were~$90.24$ and~$11.43$ seconds, respectively.
The elapsed time of the 2-D~RARMA model
is higher than the Gaussian model,
as also observed in other type of non-Gaussian models
which require iterative optimization method for estimation.

\begin{table}
	\centering
	\caption{Measures of quality
		of the fitted
		CARABAS II SAR image
		based on 2-D~RARMA$(1,1)$
		and
		2-D~ARMA$(1,1)$
		models}
	\label{t:mse2}
	\begin{tabular}{ccc}
		\toprule
		& \multicolumn{2}{c}{Model} \\
		\midrule
		& 2-D RARMA$(1,1)$ & 2-D ARMA$(1,1)$ \\
		\midrule
		MSE & $0.0562$ & $0.1241$ \\
		MAPE &  $0.4277$ & $0.7499$ \\
		\bottomrule

	\end{tabular}
\end{table}

\subsubsection{San Francisco Bay}

\begin{figure}
	\centering
	\includegraphics[scale=0.28]{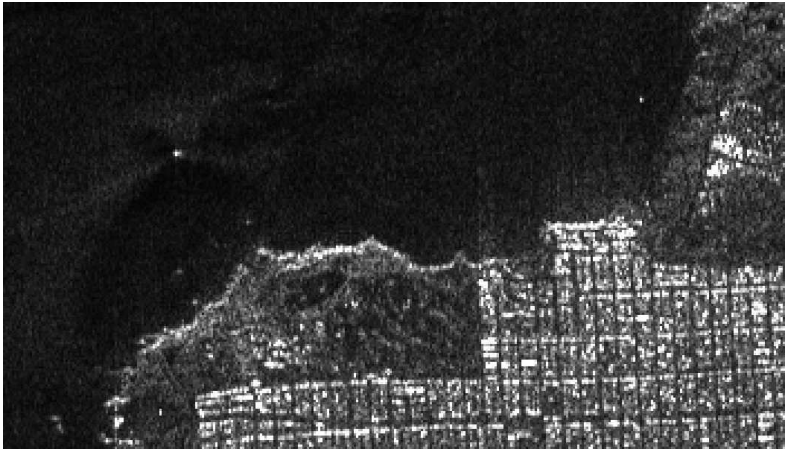}
	\caption{Original San Francisco SAR image
		HH associated polarization channel.}
	\label{f:sf_1}
\end{figure}

The considered
SAR image in this section
is the
San Francisco Bay,
obtained by the AIRSAR
sensor at band~L with four looks~\citep{Cintra2013}.
As reported in~\cite{safaee2019class},
the
San Francisco Bay image
is multilooked by a factor of
approximately
$10~\text{m}$ in azimuth
and
range.
Figure~\ref{f:sf_1}
shows
the
amplitude data
of the~$200 \times 350$
San Francisco Bay image
associated
to the
HH
polarization channel.
The ground scene of the
evaluated image
is dominated by
ocean (dark ground--top and left part of the image),
forest (gray
ground),
and urban area
(light ground--bottom)~\citep{nascimento2013}.

To perform
the ground type change detection
in the San Francisco SAR image,
we set the following parameters
in
Algorithm~\ref{a:alg1}
adopting the same methodology
described
in the previous subsection.
As discussed in~\cite{yue2019generalized},
each statistical model
is suitable for one type of specific scenario of terrain surface.
In particular, the Rayleigh distribution
is adequate  to represent homogeneous areas~\citep{oliver2004},
such as pasture, crops, and sea ground type~\citep{Cintra2013};
consequently,
the region of
interest in this section
is linked to
the ocean area.
Hence,
non-ocean regions
(forest and urban areas)
are expected
to trigger a detection,
suggesting an
anomaly
change.
Considering the methodology described in
Section~\ref{s:modelselection},
we have that
(i)~$p=1$ and $q=0$;
and
(ii)~$N=M=40$.
In the post-processing
step,
we considered
two
mathematical morphology
steps,
namely,
closing
and
opening
operations.
The dilation considered
in both steps
used
a
$10 \times  10~\text{pixels}$
square structuring element,
whose size is determined by the system resolution
and the
size of the
evaluated areas.

\begin{figure}

	\centering

	\subfigure[Northwest model]{
	\includegraphics[scale=0.62]{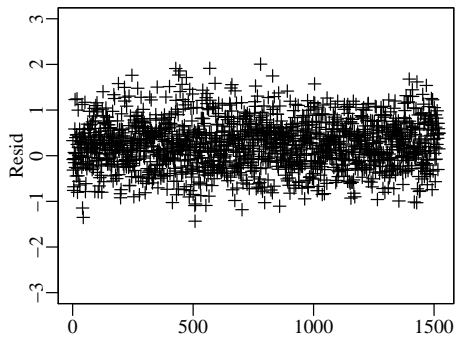}
}
\subfigure[Southwest model]{
	\includegraphics[scale=0.62]{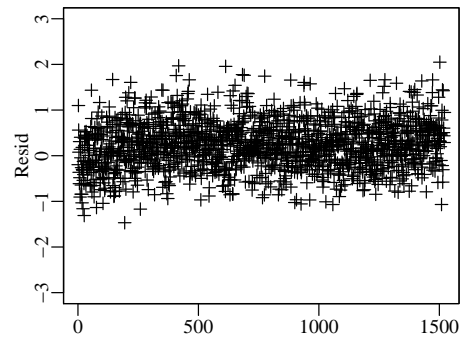}
}
\subfigure[Southeast model]{
	\includegraphics[scale=0.62]{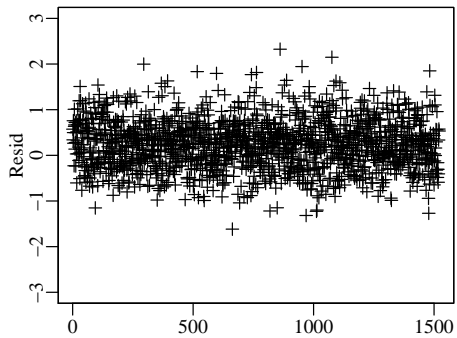}
}
\subfigure[Northeast model]{
	\includegraphics[scale=0.62]{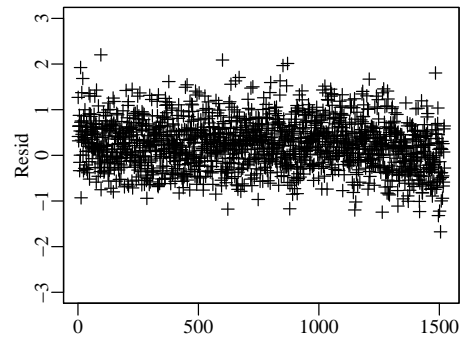}
}
	\caption{
		Residual charts of the
		2-D~RARMA$(1,0)$ models
		for the San Francisco SAR image.
	}
	\label{f:sf_hist}
\end{figure}

Because~$q=0$,
we have that~$\bm{\theta} = \mathbf{0}$,
and therefore,~$\bm{\gamma}^\star =
(\bm{\phi}^\top, \mathbf{0}^\top)^\top$
in~\eqref{e:hip_apli}.
The estimated parameters
and~SE
for each rotated image,
as described in
the second step
of the Algorithm~\ref{a:alg1},
are shown in
Table~\ref{t:sf_22}.
The $p$-values
of the Wald test are also reported in
Table~\ref{t:sf_22},
indicating
that
the spatial autocorrelation
is significant
for a probability of false alarm
equal to $0.05$.
Regarding the
diagnostic analysis of the fitted model,
as displayed in Figure~\ref{f:sf_hist},
the model residuals
are randomly distributed
and
present values close to zero.

\begin{table*}
	\centering
	\caption{
		Estimated parameters,
		standard error~(SE),
		and $p$-values
		of the
		overall significance Wald test of the
		2-D~RARMA$(1,0)$ model
		for the San Francisco SAR image}
	\label{t:sf_22}
	\begin{tabular}{cccccccccc}
		\toprule
		\multicolumn{9}{c}{Rotated Image} \\
		\midrule
			& \multicolumn{2}{c}{ Northwest}  & \multicolumn{2}{c}{ Southwest}  &
	\multicolumn{2}{c}{ Southeast} & \multicolumn{2}{c}{ Northeast}  \\
		\midrule
		& Estimate & SE & Estimate &  SE &
		Estimate &  SE & Estimate &  SE \\
						\midrule
		$\widehat{\beta}$ & $-1.2078$ &  $0.1378$ &  $-1.1879$ &  $0.1391$ & $-1.2109$ & $0.1485$  & $-1.1095$  & $0.14834$\\
		$\widehat{\phi}_{(0,1)}$ & $0.1408$ &  $0.0440$ &  $ 0.4388$ &  $0.0387$ &$ 0.1516$ &  $0.0432$ &  $ 0.4350$ &  $0.0404$\\
		$\widehat{\phi}_{(1,0)}$ & $ 0.4412$ &  $0.0389$ &  $ 0.1236$ & $0.0439$ & $ 0.4389$ &  $0.0404$ & $  0.1608$ &  $0.0431$\\
		$\widehat{\phi}_{(1,1)}$  &  $ -0.0023$ &  $0.0438$ &   $ 0.0247$  & $0.0438$ & $-0.0071$  &  $0.0452$ &$ 0.0238$ &  $0.0452$ \\
		\midrule
        $p$-value & \multicolumn{2}{c}{ $ <0.001$} &  \multicolumn{2}{c}{ $ <0.001$}
      & \multicolumn{2}{c}{ $ <0.001$} & \multicolumn{2}{c}{ $ <0.001$}  \\
		\bottomrule
	\end{tabular}
\end{table*}

\begin{figure}

	\centering

	\subfigure[2-D~RARMA$(1,0)$ model]{
		\includegraphics[scale=0.2]{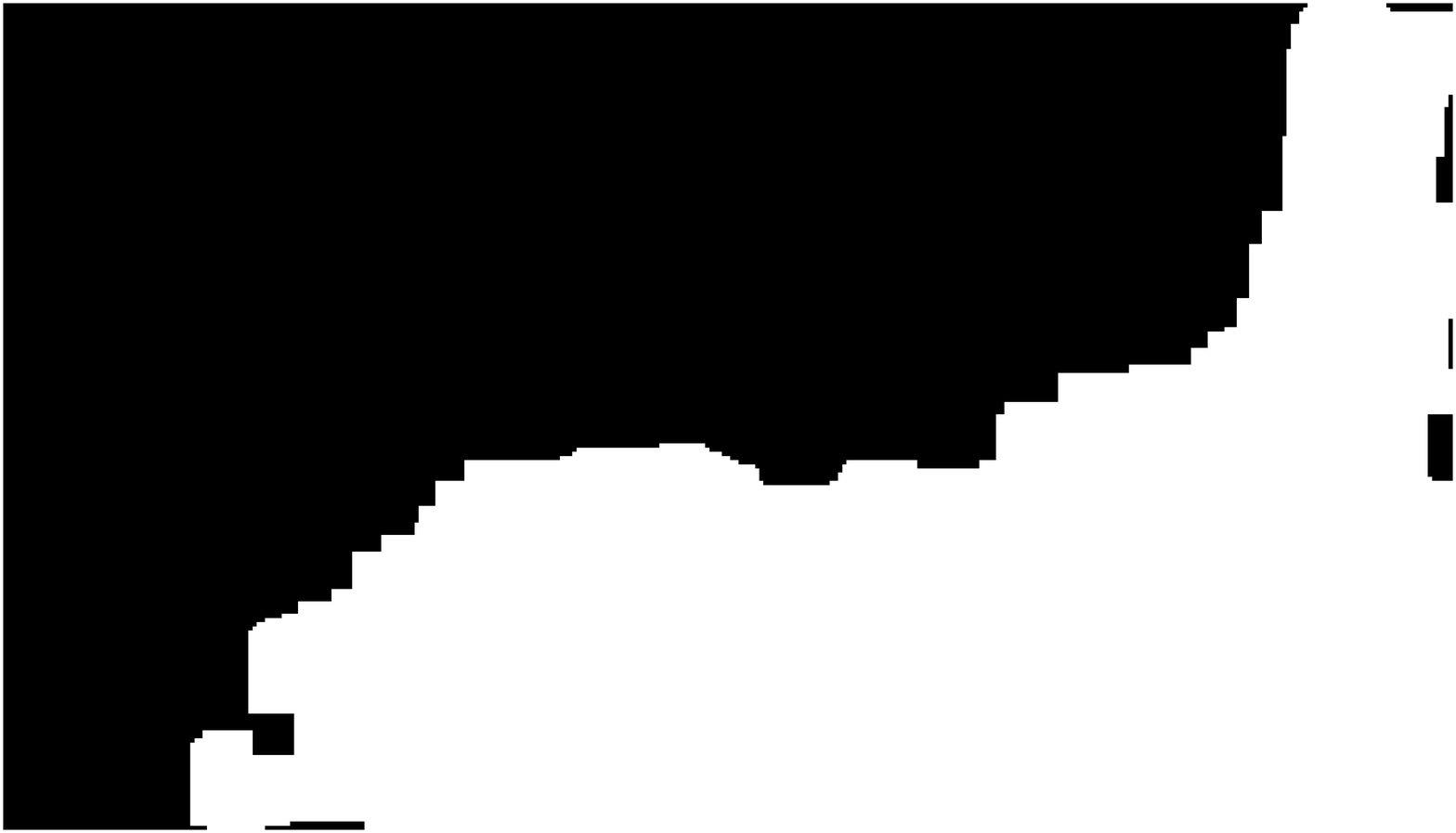}
		\label{f:detec_rar}
	}
	\subfigure[2-D~ARMA$(1,0)$ model]{
		\includegraphics[scale=0.2]{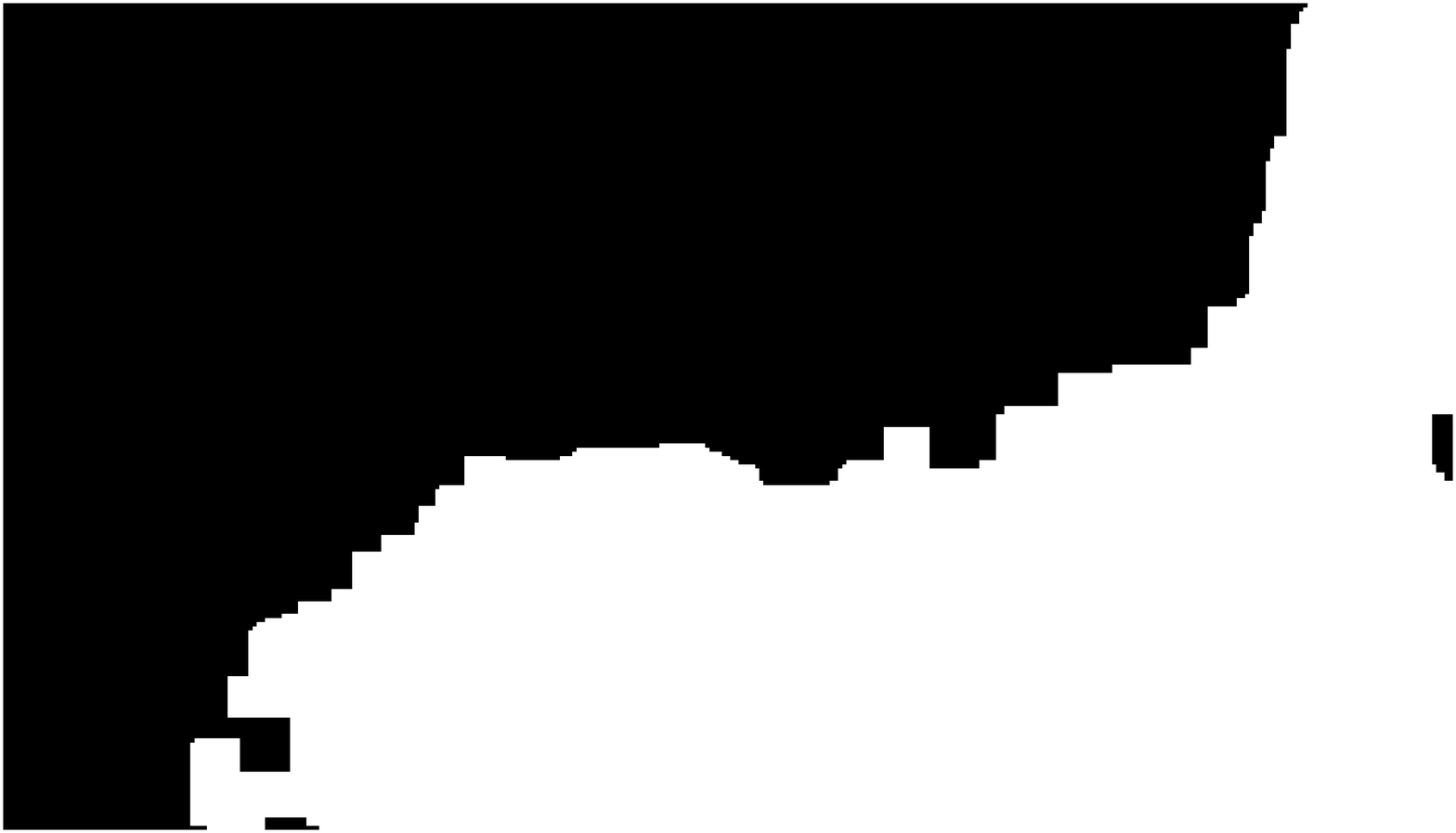}
		\label{f:detec_ar}
	}
	\caption{
		Detected images
		associated
		to the HH
		polarization channel
		based on
		the
		2-D~RARMA$(1,0)$ and 2-D~ARMA$(1,0)$ models.}
	\label{f:sf_2}
\end{figure}

The
anomaly
detection
results can be found
in Figure~\ref{f:sf_2}.
Detection
results
were compared
to the ones based
on
the
2-D~ARMA$(1,0)$.
The  results
for the detectors
considering
the
2-D~RARMA and 2-D~ARMA
can be found in
Figures~\ref{f:detec_rar}
and~\ref{f:detec_ar},
respectively.
Both evaluated detectors
identified the difference
among the ocean ground type
from the urban and forest
areas
in the tested SAR image.
We also
compared the proposed methodology
with the ones presented in~\cite{gomez2017fully},
where
machine learning optimization strategies
were used as classification techniques.
With a visual inspection,
our model and~\cite{gomez2017fully}
equally well classified
the ocean ground type.
Finally,
we computed the MSE and MAPE
figures of merit
to evaluate
the fitted image.
The 2-D~RARMA$(1,0)$
model
outperforms
the alternative model
in term of
MSE and MAPE measures,
as can be verified in
Table~\ref{t:mse1}.
The
elapsed time execution
for the 2-D~RARMA model
was equal to~$3.29$ seconds,
while for the Gaussian-based model,
we had an elapsed time of~$2.76$ seconds,
i.e.,
difference between both models
is less than one second in this experiment.

\begin{table}
	\centering
	\caption{Measures of quality
		of the fitted
		San Francisco SAR image
		based on 2-D~RARMA$(1,0)$
		and
		2-D~ARMA$(1,0)$
		models}
	\label{t:mse1}
	\begin{tabular}{ccc}
		\toprule
		& \multicolumn{2}{c}{Model} \\
		\midrule
		& 2-D RARMA$(1,0)$ & 2-D ARMA$(1,0)$ \\
		\midrule
		MSE & $0.2255$ & $ 0.3191$ \\
		MAPE &  $0.3405$ & $ 0.9711$  \\
\bottomrule

	\end{tabular}
\end{table}

\section{Conclusions}
\label{s:conclu_2d}

In this paper,
we proposed the 2-D~RARMA model
for SAR image modeling.
We introduced an inference approach for the model parameters,
the conditional Fisher information matrix,
and the asymptotic properties of the CMLE.
Monte Carlo simulations were used
to evaluate
the performance of the CMLE.
The proposed model
was applied
for
modeling and
anomaly
detection
in SAR images,
showing competitive
results when compared to
2-D~ARMA models.
Moreover,
although the proposed approach requires
much less
information
when compared to~\cite{Lundberg2006},~\cite{vu2018},
and~\cite{Vu2017},
it offered
detection results
very close
to the figures reported in~\cite{Lundberg2006},~\cite{vu2018},
and~\cite{Vu2017}
for the CARABAS~II SAR image
and
in~\cite{gomez2017fully},
for the San Francisco image.
The proposed model is presented as a suitable tool for
image modeling and
anomaly
detection
in the context
of Rayleigh distributed data,
in general,
and
SAR image processing,
in particular.

\section*{Acknowledgement}
This work was supported in part by
Conselho Nacional de Desenvolvimento
Cient\'{\i}fico e Tecnol\'ogico (CNPq),
Coordena\c{c}\~ao de Aperfei\c{c}oamento
de Pessoal de N\'{\i}vel Superior (CAPES),
Funda\c{c}\~ao de Amparo \`a Pesquisa do Estado do Rio Grande do Sul (FAPERGS),
Brazil;
and
Swedish-Brazilian Research
and Innovation Centre~(CISB),
and Saab AB, Sweden.

\appendix
\section{Conditional
	Fisher
	Information
	Matrix}

In this Appendix,
we
provide the
conditional
Fisher
information
matrix,
which is
given by
the
expectation of the negative value of the second-order
partial derivatives of the log-likelihood function,
which
is
defined as follows
\begin{align*}
\begin{split}
\dfrac{\partial^2\ell(\bm{\gamma})}{\partial\gamma_i\partial\gamma_j}
&=
\sum_{n=1}^{N}
\sum_{m=1}^{M}
\dfrac{\operatorname{ d } }{\operatorname{ d } \mu[n,m]}
\left( \dfrac{\operatorname{ d }  \ell[n,m](\mu[n,m])}{\operatorname{ d } \mu[n,m]}
\dfrac{\operatorname{ d } \mu[n,m]}{\operatorname{ d } \eta[n,m]} \right)
\dfrac{\operatorname{ d } \mu[n,m]}{\operatorname{ d } \eta[n,m]}
\dfrac{\partial\eta[n,m]}{\partial\gamma_j}
\dfrac{\partial\eta[n,m]}{\partial\gamma_i}
\\
&=
\sum_{n=1}^{N}
\sum_{m=1}^{M}
\left(
\dfrac{\partial^2 \ell[n,m](\mu[n,m])}{\partial\mu[n,m]^2}
\dfrac{\operatorname{ d }  \mu[n,m]}{\operatorname{ d }  \eta[n,m]}
+ \dfrac{\operatorname{ d }  \ell[n,m](\mu[n])}{\operatorname{ d } \mu[n,m]}
\dfrac{\partial}{\partial \mu[n,m]}
\dfrac{\operatorname{ d } \mu[n,m]}{\operatorname{ d } \eta[n,m]}
\right)
\\&
\cdot
\dfrac{\operatorname{ d } \mu[n,m]}{\operatorname{ d } \eta[n,m]}
\dfrac{\partial\eta[n,m]}{\partial\gamma_j}
\dfrac{\partial\eta[n,m]}{\partial\gamma_i}
.
\end{split}
\end{align*}
As shown in~\cite{palm2019},
we have that
$
\operatorname{E}\left( \operatorname{ d }
\ell[n,m](\mu[n,m])/\operatorname{ d }  \mu[n,m]
\mid S[n,m])
\right)=0
$.
Thus,
\begin{align*}
\begin{split}
\operatorname{E}\left(
\dfrac{\partial^2\ell(\bm{\gamma})}{\partial\gamma_i\partial\gamma_j}
\middle\vert
S[n,m])
\right)
=&
\sum_{n=1}^{N}
\sum_{m=1}^{M}
\operatorname{E}
\left(
\dfrac{\partial^2 \ell[n,m](\mu[n,m])}{\partial\mu[n,m]^2}
\right)
\left(
\dfrac{\operatorname{ d }  \mu[n,m]}{\operatorname{ d }  \eta[n,m]}
\right)^2
\dfrac{\partial\eta[n,m]}{\partial\gamma_j}
\dfrac{\partial\eta[n,m]}{\partial\gamma_i}
.
\end{split}
\end{align*}
The derivatives of
$\operatorname{ d }  \ell[n,m](\mu[n,m])/\operatorname{ d }  \mu[n,m] $,
$ \operatorname{ d }  \mu[n,m] / \operatorname{ d }  \eta[n,m]$,
and
$\partial\eta[n,m] / \partial\bm{\gamma}$
have been defined in Section~\ref{s:estimation_2d}.
Now,
the second derivative of
$\operatorname{ d }  \ell[n,m](\mu[n,m])/\operatorname{ d }  \mu[n,m] $
is given by
\begin{align*}
\frac{\partial^2 \ell[n,m](\mu[n,m])}{\partial \mu[n,m]^2}
= \frac{2}{\mu[n,m]^2} - \frac{3 \pi y[n,m]^2}{2 \mu[n,m]^4}
.
\end{align*}
Taking the expected value, we have
$
\operatorname{E}\left[\frac{\operatorname{ d } ^2
	\ell[n,m](\mu[n,m])}{\operatorname{ d }  \mu[n,m]^2} \right]
= -\frac{4}{\mu[n,m]^2}
$.
Thus,
\begin{align*}
\begin{split}
\operatorname{E}
\left[
\frac{\partial^2 \ell(\bm{\gamma})}
{\partial \gamma_i \partial \gamma_j}
\right]
=&
\sum_{n=1}^{N}
\sum_{m=1}^{M}
-\frac{4}{\mu[n,m]^2}
\left(
\frac{1}{g^\prime (\mu[n,m]}
\right)^2
\dfrac{\partial\eta[n,m]}{\partial\gamma_j}
\dfrac{\partial\eta[n,m]}{\partial\gamma_i}
.
\end{split}
\end{align*}
The conditional Fisher information matrix
is given by
\begin{align*}
\mathbf{I}(\bm{\gamma}) =
-
\setlength\arraycolsep{1pt}
\left[ \begin{array}{ccc}
I_{(\beta,\beta)} &
\mathbf{I}_{(\beta,\phi)} &
\mathbf{I}_{(\beta,\theta)}
\\
\mathbf{I}_{(\phi,\beta)} &
\mathbf{I}_{(\phi,\phi)} &
\mathbf{I}_{(\phi,\theta)}
\\
\mathbf{I}_{(\theta,\beta)} &
\mathbf{I}_{(\theta,\phi)} &
\mathbf{I}_{(\theta,\theta)}
\\
\end{array} \right]
,
\end{align*}
where
$ I_{(\beta,\beta)} = \mathbf{a^\top W a}$,
$ \mathbf{I}_{(\beta,\phi)} =
\mathbf{P^\top W a}  $,
$ \mathbf{I}_{(\beta,\theta)} =
\mathbf{R^\top W a}  $,
$ \mathbf{I}_{(\phi,\beta)} =
\mathbf{a^\top W P}  $,
$ \mathbf{I}_{(\phi,\phi)} =
\mathbf{P^\top W P}  $,
$ \mathbf{I}_{(\phi,\theta)} =
\mathbf{R^\top W P}  $,
$ \mathbf{I}_{(\theta,\alpha)} =
\mathbf{a^\top W R}  $,
$ \mathbf{I}_{(\theta,\phi)} =
\mathbf{P^\top W R}  $,
$ \mathbf{I}_{(\theta,\theta)} =
\mathbf{R^\top W R}  $.
The
matrices~$\mathbf{P}$
and~$\mathbf{R}$
are of dimensions~$(N\cdot M - w) \times (p+1)^2 -1$,
and~$(N\cdot M - w) \times (q+1)^2 -1$, respectively,
with~$(i,j)$th element
given by
$\mathbf{P}[i,j] = \frac{\partial \eta [i+w,j+w]}{\partial \phi_{(i,j)}}
$
and
$\mathbf{R}[i,j] = \frac{\partial \eta [i+w,j+w]}{\partial \theta_{(i,j)}}
$.
Finally,
we have that
\begin{align*}
\mathbf{W}
=&
\operatorname{diag}\left\{
\frac{4}{\mu[1,1]^2}\left(\dfrac{d\mu[1,1]}{d\eta[1,1]}\right)^2,
\frac{4}{\mu[1,2]^2}\left(\dfrac{d\mu[1,2]}{d\eta[1,2]}\right)^2,
\ldots,
\frac{4}{\mu[N,M]^2}\left(\dfrac{d\mu[N,M]}{d\eta[N,M]}\right)^2
\right\}
,
\end{align*}
and
$\mathbf{a} = \left( \frac{\partial \eta[n+1,m+1]}{\partial \beta},
\frac{\partial \eta[n+2,m+2]}{\partial \beta},
\ldots
,
\frac{\partial \eta[N,M]}{\partial \beta}
\right)^\top$.

{\small
\singlespacing
\bibliographystyle{unsrtnat}
\bibliography{betareg}

\begin{thebibliography}{78}
\providecommand{\natexlab}[1]{#1}
\providecommand{\url}[1]{\texttt{#1}}
\expandafter\ifx\csname urlstyle\endcsname\relax
  \providecommand{\doi}[1]{doi: #1}\else
  \providecommand{\doi}{doi: \begingroup \urlstyle{rm}\Url}\fi

\bibitem[Bustos et~al.(2009{\natexlab{a}})Bustos, Ojeda, and
  Vallejos]{Bustos2009a}
Oscar Bustos, Silvia Ojeda, and Ronny Vallejos.
\newblock Spatial {ARMA} models and its applications to image filtering.
\newblock \emph{Brazilian Journal of Probability and Statistics}, 23\penalty0
  (2):\penalty0 141--165, 2009{\natexlab{a}}.

\bibitem[Kizilkaya and Kayran(2005)]{kizilkaya2005}
Aydin Kizilkaya and A~H Kayran.
\newblock {ARMA}-cepstrum recursion algorithm for the estimation of the {MA}
  parameters of {2-D ARMA} models.
\newblock \emph{Multidimensional Systems and Signal Processing}, 16\penalty0
  (4):\penalty0 397--415, 2005.

\bibitem[Ojeda et~al.(2010)Ojeda, Vallejos, and Bustos]{ojeda2010new}
Silvia Ojeda, Ronny Vallejos, and Oscar Bustos.
\newblock A new image segmentation algorithm with applications to image
  inpainting.
\newblock \emph{Computational Statistics \& Data Analysis}, 54\penalty0
  (9):\penalty0 2082--2093, 2010.

\bibitem[Bustos et~al.(2009{\natexlab{b}})Bustos, Ruiz, Ojeda, Vallejos, and
  Frery]{bustos2009}
Oscar~H Bustos, Marcelo Ruiz, Silvia Ojeda, Ronny Vallejos, and Alejandro~C
  Frery.
\newblock Asymptotic behavior of {RA}-estimates in autoregressive {2D}
  processes.
\newblock \emph{Journal of Statistical Planning and Inference}, 139\penalty0
  (10):\penalty0 3649--3664, 2009{\natexlab{b}}.

\bibitem[Rosenfeld(2014)]{rosenfeld2014}
Azriel Rosenfeld.
\newblock \emph{Image modeling}.
\newblock Academic Press, New York, USA, 2014.

\bibitem[Nijim et~al.(1996)Nijim, Stearns, and Mikhael]{nijim1996}
Yousef~W Nijim, Samuel~D Stearns, and Wasfy~B Mikhael.
\newblock Lossless compression of images employing a linear {IIR} model.
\newblock In \emph{IEEE International Symposium on Circuits and Systems},
  volume~2, pages 305--308, 1996.

\bibitem[Chung and Kanefsky(1992)]{chung19922}
Y-S Chung and Morton Kanefsky.
\newblock On {2-D} recursive {LMS} algorithms using {ARMA} prediction for
  {ADPCM} encoding of images.
\newblock \emph{IEEE Transactions on Image Processing}, 1\penalty0
  (3):\penalty0 416--422, 1992.

\bibitem[Lim(1990)]{lim1990}
Jae~S Lim.
\newblock \emph{Two-dimensional signal and image processing}.
\newblock Prentice Hall, New Jersey, USA, 1990.

\bibitem[Vallejos and Mardesic(2004)]{vallejos2004}
Ronny~O Vallejos and Tomislav~J Mardesic.
\newblock A recursive algorithm to restore images based on robust estimation of
  {NSHP} autoregressive models.
\newblock \emph{Journal of Computational and Graphical Statistics}, 13\penalty0
  (3):\penalty0 674--682, 2004.

\bibitem[Brockwell and Davis(2016)]{brockwell2016}
Peter~J Brockwell and Richard~A Davis.
\newblock \emph{Introduction to time series and forecasting}.
\newblock Springer, Switzerland, 2016.

\bibitem[Hall and Giannakis(1995)]{hall1995}
Thomas~E Hall and Georgios~B Giannakis.
\newblock Bispectral analysis and model validation of texture images.
\newblock \emph{IEEE Transactions on Image Processing}, 4\penalty0
  (7):\penalty0 996--1009, 1995.

\bibitem[Kashyap and Eom(1988)]{kashyap1988}
Rangasami~L Kashyap and K-B Eom.
\newblock Robust image modeling techniques with an image restoration
  application.
\newblock \emph{IEEE Transactions on Acoustics, Speech, and Signal Processing},
  36\penalty0 (8):\penalty0 1313--1325, 1988.

\bibitem[Bennett and Khotanzad(1999)]{bennett1999}
Jesse Bennett and Alireza Khotanzad.
\newblock Maximum likelihood estimation methods for multispectral random field
  image models.
\newblock \emph{IEEE Transactions on Pattern Analysis and Machine
  Intelligence}, 21\penalty0 (6):\penalty0 537--543, 1999.

\bibitem[Basu and Reinsel(1993)]{basu1993}
Sabyasachi Basu and Gregory~C Reinsel.
\newblock Properties of the spatial unilateral first-order {ARMA} model.
\newblock \emph{Advances in Applied Probability}, 25\penalty0 (3):\penalty0
  631--648, 1993.

\bibitem[Cadzow and Ogino(1981)]{cadzow1981}
J~Cadzow and KOJI Ogino.
\newblock Two-dimensional spectral estimation.
\newblock \emph{IEEE Transactions on Acoustics, Speech, and Signal Processing},
  29\penalty0 (3):\penalty0 396--401, 1981.

\bibitem[Zhang(1991)]{zhang1991}
X-D Zhang.
\newblock On the estimation of two-dimensional moving average parameters.
\newblock \emph{IEEE Transactions on Automatic Control}, 36\penalty0
  (10):\penalty0 1196--1199, 1991.

\bibitem[Zoubir et~al.(2018)Zoubir, Koivunen, Ollila, and Muma]{zoubir2018}
Abdelhak~M Zoubir, Visa Koivunen, Esa Ollila, and Michael Muma.
\newblock \emph{Robust statistics for signal processing}.
\newblock Cambridge University Press, New York, USA, 2018.

\bibitem[Wegman et~al.(1989)Wegman, Schwartz, and Thomas]{schwartz1989topics}
Edward~J Wegman, Stuart~C Schwartz, and John~Bowman Thomas.
\newblock \emph{Topics in non-{G}aussian signal processing}.
\newblock Springer, New York, USA, 1989.

\bibitem[Zhao et~al.(2008)Zhao, Popescu, and Zhang]{zhao2008}
Kaiguang Zhao, Sorin Popescu, and Xuesong Zhang.
\newblock Bayesian learning with {G}aussian processes for supervised
  classification of hyperspectral data.
\newblock \emph{Photogrammetric Engineering \& Remote Sensing}, 74\penalty0
  (10):\penalty0 1223--1234, 2008.

\bibitem[Zhao et~al.(2016)Zhao, Zhong, Ma, and Zhang]{zhao2016}
Bei Zhao, Yanfei Zhong, Ailong Ma, and Liangpei Zhang.
\newblock A spatial {G}aussian mixture model for optical remote sensing image
  clustering.
\newblock \emph{IEEE Journal of Selected Topics in Applied Earth Observations
  and Remote Sensing}, 9\penalty0 (12):\penalty0 5748--5759, 2016.

\bibitem[Morales-Alvarez et~al.(2017)Morales-Alvarez, P{\'e}rez-Suay, Molina,
  and Camps-Valls]{morales2017}
Pablo Morales-Alvarez, Adri{\'a}n P{\'e}rez-Suay, Rafael Molina, and Gustau
  Camps-Valls.
\newblock Remote sensing image classification with large-scale {G}aussian
  processes.
\newblock \emph{IEEE Transactions on Geoscience and Remote Sensing},
  56\penalty0 (2):\penalty0 1103--1114, 2017.

\bibitem[Kay(2000)]{kay2000}
Steven~M Kay.
\newblock Can detectability be improved by adding noise?
\newblock \emph{IEEE Signal Processing Letters}, 7\penalty0 (1):\penalty0
  8--10, 2000.

\bibitem[Xue et~al.(2020)Xue, Xu, and Shui]{xue2020}
Jian Xue, Shuwen Xu, and Penglang Shui.
\newblock Near-optimum coherent {CFAR} detection of radar targets in
  compound-{G}aussian clutter with inverse {G}aussian texture.
\newblock \emph{Signal Processing}, 166:\penalty0 107236, 2020.

\bibitem[Wang et~al.(2019)Wang, Lopez-Molina, de~Ulzurrun, and
  De~Baets]{wang2019}
Gang Wang, Carlos Lopez-Molina, Guillermo Vidal-Diez de~Ulzurrun, and Bernard
  De~Baets.
\newblock Noise-robust line detection using normalized and adaptive
  second-order anisotropic {G}aussian kernels.
\newblock \emph{Signal Processing}, 160:\penalty0 252--262, 2019.

\bibitem[Margoosian et~al.(2015)Margoosian, Abouei, and
  Plataniotis]{margoosian2015}
Argin Margoosian, Jamshid Abouei, and Konstantinos~N Plataniotis.
\newblock An accurate kernelized energy detection in {G}aussian and
  non-{G}aussian/impulsive noises.
\newblock \emph{IEEE Transactions on Signal Processing}, 63\penalty0
  (21):\penalty0 5621--5636, 2015.

\bibitem[Liu et~al.(2018)Liu, Liu, Liu, Zhou, Zhu, and Zhang]{liu2018}
Jun Liu, Sha Liu, Weijian Liu, Shenghua Zhou, Shengqi Zhu, and Zi-Jing Zhang.
\newblock Persymmetric adaptive detection of distributed targets in
  compound-{G}aussian sea clutter with {G}amma texture.
\newblock \emph{Signal Processing}, 152:\penalty0 340--349, 2018.

\bibitem[Bayer et~al.(2020{\natexlab{a}})Bayer, Bayer, Marinoni, and
  Gamba]{Bayer2019b}
F.~M. Bayer, D.~M. Bayer, A.~Marinoni, and P.~Gamba.
\newblock A novel {R}ayleigh dynamical model for remote sensing data
  interpretation.
\newblock \emph{IEEE Transactions on Geoscience and Remote Sensing},
  58\penalty0 (7):\penalty0 4989 -- 4999, 2020{\natexlab{a}}.

\bibitem[Zanetti et~al.(2015)Zanetti, Bovolo, and Bruzzone]{zanetti2015}
Massimo Zanetti, Francesca Bovolo, and Lorenzo Bruzzone.
\newblock Rayleigh-rice mixture parameter estimation via {EM} algorithm for
  change detection in multispectral images.
\newblock \emph{IEEE Transactions on Image Processing}, 24\penalty0
  (12):\penalty0 5004--5016, 2015.

\bibitem[Sumaiya and Kumari(2018)]{sumaiya2018}
Mohammed~Naina Sumaiya and Ramapackiam Shantha~Selva Kumari.
\newblock Unsupervised change detection of flood affected areas in {SAR} images
  using {R}ayleigh-based {B}ayesian thresholding.
\newblock \emph{IET Radar, Sonar \& Navigation}, 12\penalty0 (5):\penalty0
  515--522, 2018.

\bibitem[Yue et~al.(2019)Yue, Xu, Frery, and Jin]{yue2019generalized}
Dong-Xiao Yue, Feng Xu, Alejandro~C Frery, and Ya-Qiu Jin.
\newblock A generalized {G}aussian coherent scatterer model for correlated
  {SAR} texture.
\newblock \emph{IEEE Transactions on Geoscience and Remote Sensing},
  58\penalty0 (4), 2019.

\bibitem[Kuruoglu and Zerubia(2004)]{Kuruoglu2004}
Ercan~E Kuruoglu and Josiane Zerubia.
\newblock Modeling {SAR} images with a generalization of the {R}ayleigh
  distribution.
\newblock \emph{IEEE Transactions on Image Processing}, 13\penalty0
  (4):\penalty0 527--533, 2004.

\bibitem[Jackson and Moses(2009)]{Jackson2009}
Julie~Ann Jackson and Randolph~L Moses.
\newblock A model for generating synthetic {VHF} {SAR} forest clutter images.
\newblock \emph{IEEE Transactions on Aerospace and Electronic Systems},
  45\penalty0 (3):\penalty0 1138 -- 1152, 2009.

\bibitem[Kuttikkad and Chellappa(2000)]{kuttikkad2000statistical}
Shyam Kuttikkad and Rama Chellappa.
\newblock Statistical modeling and analysis of high-resolution synthetic
  aperture radar images.
\newblock \emph{Statistics and Computing}, 10\penalty0 (2):\penalty0 133--145,
  2000.

\bibitem[Oliver and Quegan(2004)]{oliver2004}
Chris Oliver and Shaun Quegan.
\newblock \emph{Understanding synthetic aperture radar images}.
\newblock SciTech Publishing, USA, 2004.

\bibitem[Yue et~al.(2021)Yue, Xu, Frery, and Jin]{Yue2021}
Dong-Xiao Yue, Feng Xu, Alejandro~C. Frery, and Ya-Qiu Jin.
\newblock Synthetic aperture radar image statistical modeling: Part
  one-single-pixel statistical models.
\newblock \emph{IEEE Geoscience and Remote Sensing Magazine}, 9\penalty0
  (1):\penalty0 82--114, 2021.

\bibitem[Wang and Ouchi(2008)]{wang2008}
Haipeng Wang and Kazuo Ouchi.
\newblock Accuracy of the $ k $-distribution regression model for forest
  biomass estimation by high-resolution polarimetric {SAR}: Comparison of model
  estimation and field data.
\newblock \emph{IEEE Transactions on Geoscience and Remote Sensing},
  46\penalty0 (4):\penalty0 1058--1064, 2008.

\bibitem[McCullagh and Nelder(1989)]{McCullagh1989}
P.~McCullagh and J.A. Nelder.
\newblock \emph{Generalized linear models}.
\newblock Chapman and Hall, New York, USA, 2nd edition, 1989.

\bibitem[Palm et~al.(2019)Palm, Bayer, Cintra, Pettersson, and
  Machado]{palm2019}
Bruna~G Palm, F{\'a}bio~M Bayer, Renato~J Cintra, Mats~I Pettersson, and Renato
  Machado.
\newblock Rayleigh regression model for ground type detection in {SAR} imagery.
\newblock \emph{IEEE Geoscience and Remote Sensing Letters}, 16\penalty0
  (10):\penalty0 1660 -- 1664, 2019.

\bibitem[Yan et~al.(2018)Yan, Paynabar, and Shi]{yan2018real}
Hao Yan, Kamran Paynabar, and Jianjun Shi.
\newblock Real-time monitoring of high-dimensional functional data streams via
  spatio-temporal smooth sparse decomposition.
\newblock \emph{Technometrics}, 60\penalty0 (2):\penalty0 181--197, 2018.

\bibitem[Almeida-Junior and Nascimento(2021)]{almeida2021}
Pedro~M Almeida-Junior and Abra{\~a}o~DC Nascimento.
\newblock ${G}_i^0$ {ARMA} process for speckled data.
\newblock \emph{Journal of Statistical Computation and Simulation}, 2021.
\newblock \doi{10.1080/00949655.2021.1922688}.

\bibitem[Benjamin et~al.(2003)Benjamin, Rigby, and Stasinopoulos]{benjamin2003}
Michael~A Benjamin, Robert~A Rigby, and D~Mikis Stasinopoulos.
\newblock Generalized autoregressive moving average models.
\newblock \emph{Journal of the American Statistical Association}, 98\penalty0
  (461):\penalty0 214--223, 2003.

\bibitem[Bayer et~al.(2017)Bayer, Bayer, and Pumi]{bayer2017}
F{\'a}bio~Mariano Bayer, D{\'e}bora~Missio Bayer, and Guilherme Pumi.
\newblock Kumaraswamy autoregressive moving average models for double bounded
  environmental data.
\newblock \emph{Journal of Hydrology}, 555:\penalty0 385--396, 2017.

\bibitem[Rocha and Cribari-Neto(2009)]{rocha2009}
A.~V. Rocha and F.~Cribari-Neto.
\newblock Beta autoregressive moving average models.
\newblock \emph{Test}, 18\penalty0 (3):\penalty0 529--545, 2009.

\bibitem[M{\"o}ller and Wei{\ss}(2020)]{moller2020generalized}
Tobias~A M{\"o}ller and Christian~H Wei{\ss}.
\newblock Generalized discrete autoregressive moving-average models.
\newblock \emph{Applied Stochastic Models in Business and Industry},
  36\penalty0 (4):\penalty0 641--659, 2020.

\bibitem[Palm et~al.(2021)Palm, Bayer, and Cintra]{Palm2021}
Bruna~G. Palm, Fábio~M. Bayer, and Renato~J. Cintra.
\newblock Signal detection and inference based on the beta binomial
  autoregressive moving average model.
\newblock \emph{Digital Signal Processing}, 109:\penalty0 102911, 2021.

\bibitem[Bayer et~al.(2020{\natexlab{b}})Bayer, Bayer, and Gamba]{bayer20203}
D{\'e}bora~M Bayer, F{\'a}bio~M Bayer, and Paolo Gamba.
\newblock A 3-{D} spatiotemporal model for remote sensing data cubes.
\newblock \emph{IEEE Transactions on Geoscience and Remote Sensing},
  52\penalty0 (2):\penalty0 1082--1093, 2020{\natexlab{b}}.

\bibitem[Brooks et~al.(2013)Brooks, Wynne, Thomas, Blinn, and
  Coulston]{brooks2013}
Evan~B Brooks, Randolph~H Wynne, Valerie~A Thomas, Christine~E Blinn, and
  John~W Coulston.
\newblock On-the-fly massively multitemporal change detection using statistical
  quality control charts and {L}andsat data.
\newblock \emph{IEEE Transactions on Geoscience and Remote Sensing},
  52\penalty0 (6):\penalty0 3316--3332, 2013.

\bibitem[Talagala et~al.(2020)Talagala, Hyndman, and Smith-Miles]{talagala2020}
Priyanga~Dilini Talagala, Rob~J Hyndman, and Kate Smith-Miles.
\newblock Anomaly detection in high-dimensional data.
\newblock \emph{Journal of Computational and Graphical Statistics}, 30\penalty0
  (2):\penalty0 360--374, 2020.

\bibitem[Kadri et~al.(2016)Kadri, Harrou, Chaabane, Sun, and Tahon]{kadri2016}
Farid Kadri, Fouzi Harrou, Sond{\`e}s Chaabane, Ying Sun, and Christian Tahon.
\newblock Seasonal {ARMA}-based {SPC} charts for anomaly detection:
  {A}pplication to emergency department systems.
\newblock \emph{Neurocomputing}, 173:\penalty0 2102--2114, 2016.

\bibitem[Quatrini et~al.(2020)Quatrini, Costantino, Di~Gravio, and
  Patriarca]{quatrini2020}
Elena Quatrini, Francesco Costantino, Giulio Di~Gravio, and Riccardo Patriarca.
\newblock Machine learning for anomaly detection and process phase
  classification to improve safety and maintenance activities.
\newblock \emph{Journal of Manufacturing Systems}, 56:\penalty0 117--132, 2020.

\bibitem[Kwon and Nasrabadi(2005)]{kwon2005}
Heesung Kwon and Nasser~M Nasrabadi.
\newblock Kernel {RX}-algorithm: {A} nonlinear anomaly detector for
  hyperspectral imagery.
\newblock \emph{IEEE Transactions on Geoscience and Remote Sensing},
  43\penalty0 (2):\penalty0 388--397, 2005.

\bibitem[Melchior et~al.(2021)Melchior, Zanini, Guerra, and
  Rockenbach]{melchior2021}
Cristiane Melchior, Roselaine~Ruviaro Zanini, Renata~Rojas Guerra, and Dinei~A
  Rockenbach.
\newblock Forecasting {B}razilian mortality rates due to occupational accidents
  using autoregressive moving average approaches.
\newblock \emph{International Journal of Forecasting}, 37\penalty0
  (2):\penalty0 825--837, 2021.

\bibitem[Leiva et~al.(2020)Leiva, Saulo, Souza, Aykroyd, and Vila]{leiva2020}
V{\'\i}ctor Leiva, Helton Saulo, Rubens Souza, Robert~G Aykroyd, and Roberto
  Vila.
\newblock A new {BISARMA} time series model for forecasting mortality using
  weather and particulate matter data.
\newblock \emph{Journal of Forecasting}, 40\penalty0 (2):\penalty0 346--364,
  2020.

\bibitem[Rothermel et~al.(2020)Rothermel, Balazik, Best, Breece, Fox, Gahagan,
  Haulsee, Higgs, O’Brien, Oliver, et~al.]{rothermel2020}
Ella~R Rothermel, Matthew~T Balazik, Jessica~E Best, Matthew~W Breece,
  Dewayne~A Fox, Benjamin~I Gahagan, Danielle~E Haulsee, Amanda~L Higgs,
  Michael~HP O’Brien, Matthew~J Oliver, et~al.
\newblock Comparative migration ecology of striped bass and {A}tlantic sturgeon
  in the {US} {S}outhern mid-{A}tlantic bight flyway.
\newblock \emph{PloS One}, 15\penalty0 (6):\penalty0 e0234442, 2020.

\bibitem[Liboschik et~al.(2017)Liboschik, Fokianos, and Fried]{liboschik2017}
Tobias Liboschik, Konstantinos Fokianos, and Roland Fried.
\newblock tscount: {A}n {R} package for analysis of count time series following
  generalized linear models.
\newblock \emph{Journal of Statistical Software}, 82\penalty0 (1):\penalty0
  1--51, 2017.

\bibitem[Scher et~al.(2020)Scher, Cribari-Neto, Pumi, and Bayer]{scher2020}
Vin{\'\i}cius~T Scher, Francisco Cribari-Neto, Guilherme Pumi, and F{\'a}bio~M
  Bayer.
\newblock Goodness-of-fit tests for $\beta${ARMA} hydrological time series
  modeling.
\newblock \emph{Environmetrics}, 31\penalty0 (3):\penalty0 e2607, 2020.

\bibitem[Press et~al.(1992)Press, Teukolsky, Vetterling, and Flannery]{press}
W.~Press, S.~Teukolsky, W.~Vetterling, and B.~Flannery.
\newblock \emph{Numerical recipes in {C}: {T}he art of scientific computing}.
\newblock Cambridge University Press, New York, USA, 2nd edition edition, 1992.

\bibitem[Nocedal and Wright(1999)]{nocedal1999}
J.~Nocedal and S.~J. Wright.
\newblock \emph{Numerical optimization}.
\newblock Springer, New York, USA, 1999.

\bibitem[Mittelhammer et~al.(2000)Mittelhammer, Judge, and
  Miller]{Mittelhammer2000}
R.~C. Mittelhammer, G.~G. Judge, and D.~J. Miller.
\newblock \emph{Econometric Foundations}.
\newblock Cambridge University Press, Cambridge, UK, 2000.

\bibitem[Andersen(1970)]{andersen1970asymptotic}
Erling~Bernhard Andersen.
\newblock Asymptotic properties of conditional maximum-likelihood estimators.
\newblock \emph{Journal of the Royal Statistical Society: Series B
  (Methodological)}, 32\penalty0 (2):\penalty0 283--301, 1970.

\bibitem[Pawitan(2001)]{Pawitan2001}
Y.~Pawitan.
\newblock \emph{In all likelihood: {S}tatistical modelling and inference using
  likelihood}.
\newblock Oxford Science publications, UK, 2001.

\bibitem[Kay(1998)]{Kay1998-2}
Steven~M. Kay.
\newblock \emph{Fundamentals of statistical signal processing: {D}etection
  theory}, volume~II.
\newblock Prentice Hall, Upper Saddle River, NJ, USA, 1998.

\bibitem[Kedem and Fokianos(2005)]{kedem2005}
Benjamin Kedem and Konstantinos Fokianos.
\newblock \emph{Regression models for time series analysis}.
\newblock John Wiley \& Sons, New Jersey, USA, 2005.

\bibitem[Dunn and Smyth(1996)]{Dunn1996}
Peter~K. Dunn and Gordon~K. Smyth.
\newblock Randomized quantile residuals.
\newblock \emph{Journal of Computational and Graphical Statistics}, 5\penalty0
  (3):\penalty0 236--244, 1996.

\bibitem[Bayer et~al.(2019)Bayer, Kozakevicius, and Cintra]{bayer2019}
F{\'a}bio~M Bayer, Alice~J Kozakevicius, and Renato~J Cintra.
\newblock An iterative wavelet threshold for signal denoising.
\newblock \emph{Signal Processing}, 162:\penalty0 10--20, 2019.

\bibitem[Edmond et~al.(2000)Edmond, Ronald, and Hugues]{edmond2000mathematical}
J~Edmond, J~Ronald, and T~Hugues.
\newblock Mathematical morphology: {A} useful set of tools for image analysis.
\newblock \emph{Statistics and Computing}, 10\penalty0 (2):\penalty0 105--120,
  2000.

\bibitem[Gonzalez et~al.(2009)Gonzalez, Woods, and Eddine]{gonzales2009}
R~Gonzalez, R~Woods, and S~Eddine.
\newblock Digital image processing using {MATLAB}, 2009.

\bibitem[Box et~al.(2008)Box, Jenkins, and Reinsel]{Box2008}
G.~Box, G.~M. Jenkins, and G.~Reinsel.
\newblock \emph{Time series analysis: {F}orecasting and control}.
\newblock Hardcover, John Wiley \& Sons, USA, June 2008.

\bibitem[Akaike(1974)]{Akaike}
Hirotugu Akaike.
\newblock A new look at the statistical model identification.
\newblock \emph{IEEE Transactions on Automatic Control}, 19\penalty0
  (6):\penalty0 716--723, 1974.

\bibitem[Schwarz et~al.(1978)]{schwarz1978}
Gideon Schwarz et~al.
\newblock Estimating the dimension of a model.
\newblock \emph{The {A}nnals of {S}tatistics}, 6\penalty0 (2):\penalty0
  461--464, 1978.

\bibitem[Lundberg et~al.(2006)Lundberg, Ulander, Pierson, and
  Gustavsson]{Lundberg2006}
Mikael Lundberg, Lars M.~H. Ulander, William~E. Pierson, and Anders Gustavsson.
\newblock A challenge problem for detection of targets in foliage.
\newblock In \emph{Proc. SPIE}, volume 6237, page 62370K, 2006.

\bibitem[Ulander et~al.(2005)Ulander, Lundberg, Pierson, and
  Gustavsson]{Ulander2005}
Lars~MH Ulander, M~Lundberg, W~Pierson, and A~Gustavsson.
\newblock Change detection for low-frequency {SAR} ground surveillance.
\newblock \emph{IEEE Proceedings-Radar, Sonar and Navigation}, 152\penalty0
  (6):\penalty0 413--420, 2005.

\bibitem[Vu et~al.(2018)Vu, Gomes, Pettersson, Dammert, and Hellsten]{vu2018}
Viet~Thuy Vu, Natanael~Rodrigues Gomes, Mats~I Pettersson, Patrik Dammert, and
  Hans Hellsten.
\newblock Bivariate gamma distribution for wavelength-resolution {SAR} change
  detection.
\newblock \emph{IEEE Transactions on Geoscience and Remote Sensing},
  57\penalty0 (1):\penalty0 473--481, 2018.

\bibitem[Vu(2017)]{Vu2017}
Viet~Thuy Vu.
\newblock Wavelength-resolution {SAR} incoherent change detection based on
  image stack.
\newblock \emph{IEEE Geoscience and Remote Sensing Letters}, 14\penalty0
  (7):\penalty0 1012--1016, 2017.

\bibitem[Cintra et~al.(2013)Cintra, Frery, and Nascimento]{Cintra2013}
Renato~J Cintra, Alejandro~C Frery, and Abraao~DC Nascimento.
\newblock Parametric and nonparametric tests for speckled imagery.
\newblock \emph{Pattern Analysis and Applications}, 16\penalty0 (2):\penalty0
  141--161, 2013.

\bibitem[Safaee and Sahebi(2019)]{safaee2019class}
Bahram Safaee and Mahmod~Reza Sahebi.
\newblock A class-based approach to classify {P}ol{SAR} imagery using optimum
  classifier.
\newblock \emph{European Journal of Remote Sensing}, 52\penalty0 (1):\penalty0
  294--307, 2019.

\bibitem[Nascimento et~al.(2013)Nascimento, Horta, Frery, and
  Cintra]{nascimento2013}
Abra{\~a}o~DC Nascimento, Michelle~M Horta, Alejandro~C Frery, and Renato~J
  Cintra.
\newblock Comparing edge detection methods based on stochastic entropies and
  distances for {PolSAR} imagery.
\newblock \emph{IEEE Journal of Selected Topics in Applied Earth Observations
  and Remote Sensing}, 7\penalty0 (2):\penalty0 648--663, 2013.

\bibitem[Gomez et~al.(2017)Gomez, Alvarez, Mazorra, and Frery]{gomez2017fully}
Luis Gomez, Luis Alvarez, Luis Mazorra, and Alejandro~C Frery.
\newblock Fully {P}ol{SAR} image classification using machine learning
  techniques and reaction-diffusion systems.
\newblock \emph{Neurocomputing}, 255:\penalty0 52--60, 2017.

\end{thebibliography}
}

\end{document}